%% file: main.tex
\documentclass[
  english,
  acmsmall,
  authorversion,
  ]{acmart}

\usepackage{listings}
\usepackage{subcaption}

\usepackage{longtable}

\usepackage{acronym}
\input{glossary}

\usepackage{color}

\newcommand{\SI}[2]{%
  #1\,#2%
}

\input{aliases}

\lstdefinestyle{html}{
  basicstyle=\footnotesize\fontfamily{pxtt}\selectfont,
  keywordstyle=\bfseries,
  showstringspaces=false,
  frame=tb,
}

\title{\textsc{Breaking Bad}: Quantifying the Addiction of Web Elements to JavaScript}
\author{Romain Fouquet}
\orcid{0000-0003-0369-7459}
\affiliation{%
  \institution{Univ. Lille, Inria, CNRS, Centrale Lille, UMR 9189 CRIStAL}
  \postcode{F-59000}
  \city{Lille}
  \country{France}}
\email{romain.fouquet@inria.fr}
\author{Pierre Laperdrix}
\orcid{0000-0001-6901-3596}
\affiliation{%
  \institution{CNRS, Univ Lille, Inria Lille}
  \city{Lille}
  \country{France}}
\email{pierre.laperdrix@inria.fr}
\author{Romain Rouvoy}
\orcid{0000-0003-1771-8791}
\affiliation{%
  \institution{Univ.\,Lille / Inria / IUF}
  \city{Lille}
  \country{France}}
\email{romain.rouvoy@univ-lille.fr}



\setcopyright{none}
\acmJournal{TOIT}
\acmYear{2023} \acmVolume{1} \acmNumber{1} \acmArticle{1} \acmMonth{1} \acmPrice{15.00}\acmDOI{10.1145/3579846}

\date{}

\begin{document}
\input{abstract}

\maketitle

\input{introduction}

\input{background}

\input{methodology}

\input{results}

\input{limitations}

\input{discussion}

\input{conclusion}

\bibliographystyle{ACM-Reference-Format}
\bibliography{references, paper_references}

\input{appendix}
\end{document}

%% file: glossary.tex
\newacro{dom}[DOM]{Document Object Model}
\newacro{url}[URL]{Uniform Resource Locator}
\newacro{html}[HTML]{HyperText Markup Language}
\newacro{css}[CSS]{Cascading Style Sheets}
\newacro{js}[JS]{JavaScript}
\newacro{json}[JSON]{JavaScript Object Notation}
\newacro{idl}[IDL]{Interface Description Language}
\newacro{gif}[GIF]{Graphics Interchange Format}
\newacro{http}[HTTP]{HyperText Transfer Protocol}
\newacro{api}[API]{Application Programming Interface}
\newacro{ui}[UI]{User Interface}
\newacro{seo}[SEO]{Search Engine Optimization}
\newacro{waiaria}[WAI-ARIA]{Web Accessibility Initiative --- Accessible Rich Internet Applications}
\newacro{ssr}[SSR]{Server-Side Rendering}
\newacro{ssg}[SSG]{Static Site Generation}
\newacro{ajax}[AJAX]{Asynchronous JavaScript and XML}
\newacro{xhr}[XHR]{XML HTTP Request}

\newacro{dbr}[DBR]{Differential BReakage}
\newacro{dbrn}[DBRn]{Differential BReakage, Normalized}

%% file: aliases.tex
\newcommand{\na}{N/A}

\newcommand{\javascript}{Java\-Script}

\newcommand{\firefox}{Firefox}
\newcommand{\firefoxetp}{\firefox{} Enhanced Tracking Protection}
\newcommand{\hispar}{Hispar}
\newcommand{\alexatoponem}{Alexa Top~1M}
\newcommand{\google}{Google}
\newcommand{\googlesearch}{Google Search}
\newcommand{\youtube}{YouTube}
\newcommand{\twitter}{Twitter}
\newcommand{\instagram}{Instagram}
\newcommand{\puppeteer}{Puppeteer}
\newcommand{\httparchive}{HTTP Archive}

\newcommand{\disconnect}{Disconnect}
\newcommand{\opendns}{Open\-DNS}
\newcommand{\github}{GitHub}
\newcommand{\wcubetechs}{W3Techs}

\newcommand{\webextensions}{WebExtensions}

\newcommand{\bootstrap}{Bootstrap}
\newcommand{\zurbfoundation}{ZURB Foundation}
\newcommand{\semanticui}{Semantic UI}

\newcommand{\react}{React}
\newcommand{\angularjs}{AngularJS}
\newcommand{\vuejs}{Vue.JS}
\newcommand{\svelte}{Svelte}

\newcommand{\nextjs}{Next.js}
\newcommand{\nuxtjs}{NuxtJS}

\newcommand{\ublockorigin}{uBlock Origin}
\newcommand{\noscripext}{NoScript}
\newcommand{\umatrix}{uMatrix}

\newcommand{\htmltag}[1]{\texttt{<#1>}}
\newcommand{\htmlattr}[1]{\texttt{#1}}
\newcommand{\htmlattrvalue}[1]{\texttt{"#1"}} 
\newcommand{\idlattr}[1]{\texttt{#1}}

\newcommand{\event}[1]{\texttt{#1}}
\newcommand{\csspseudoclass}[1]{\texttt{:#1}}

\newcommand{\cssclass}[1]{\texttt{.#1}}

\newcommand{\firefoxpref}[1]{\texttt{#1}}

\newcommand{\domloadtimeout}{\SI{10}{s}}
\newcommand{\loadtimeout}{\SI{30}{s}}
\newcommand{\inspectiontimeout}{\SI{60}{s}}
\newcommand{\idletime}{\SI{3}{s}}

\newcommand{\brokencount}[3]{\mathrm{BrokenCount}^{#1}_{\mathrm{#2}_{\mathrm{#3}}}}
\newcommand{\workingcount}[3]{\mathrm{WorkingCount}^{#1}_{\mathrm{#2}_{\mathrm{#3}}}}
\newcommand{\dbr}[2]{\mathrm{DBR}_{\mathrm{#1}_{\mathrm{#2}}}}
\newcommand{\dbrn}[2]{\mathrm{DBRn}_{\mathrm{#1}_{\mathrm{#2}}}}
\newcommand{\totcnt}[3]{\mathrm{TotalCount}^{#1}_{\mathrm{#2}_{\mathrm{#3}}}}

\newcommand{\crawlstartdate}{2021-03-26}
\newcommand{\crawlenddate}{2021-03-28}

\newcommand{\crawlplain}{\texttt{[plain]}}
\newcommand{\crawlnojs}{\texttt{[nojs]}}

\newcommand{\largesizepx}{\SI{100}{px}}

\newcommand{\medianloadtimeplain}{\SI{3,173}{ms}}
\newcommand{\medianloadtimenojs}{\SI{1,582}{ms}}

\newcommand{\crawlurls}{\SI{6,384}{pages}}

\newcommand{\crawlurlsbothloaded}{\SI{5,827}{pages}}
\newcommand{\crawlurlsbothfeatures}{\SI{5,695}{pages}}
\newcommand{\crawlurlsbothhtml}{\SI{4,911}{pages}}
\newcommand{\crawldomains}{\SI{3,774}{domains}}
\newcommand{\crawlalexabasedomains}{\SI{2,136}{\text{Alexa base domains}}}

\newcommand{\percdomainscat}{\SI{66}{\%}}

\newcommand{\perckindamainsec}{\SI{37}{\%}}
\newcommand{\perckindasemantic}{\SI{76}{\%}}

\newcommand{\percallworking}{\SI{43}{\%}}
\newcommand{\percmainworking}{\SI{67}{\%}}
\newcommand{\percintermain}{\SI{25}{\%}}

\newcommand{\perclesstrackingnojs}{\SI{85}{\%}}

\newcommand{\catthreshold}{50~pages}

\newcommand{\crawlbrowser}{Firefox Nightly~88.0a1}

\newcommand{\emailprotected}{\texttt{[email protected]}}

\newcommand{\linefil}{\leaders\vrule height.4pt\hfill}
\newcommand{\featurebox}[2]{%
  \bigskip\hbox{\vrule \vbox{\noindent
    \linefil \quad \smash{\lower.5ex\hbox{\it User-perceived breakage symptoms}}\quad \linefil\null \par
    \nointerlineskip \vbox{\bigskip\leftskip=1.5em \rightskip=1.5em #2\bigskip}
    \hrule}\vrule}
  \medskip
}

%% file: abstract.tex
\begin{abstract}
While \javascript{} established itself as a cornerstone of the modern web, it
also constitutes a major tracking and security vector, thus raising critical
privacy and security concerns.
In this context, some browser extensions propose to systematically block scripts
reported by crowdsourced trackers lists.
However, this solution heavily depends on the quality of these built-in lists,
which may be deprecated or incomplete, thus exposing the visitor to unknown
trackers.
In this paper, we explore a different strategy, by investigating the benefits of
disabling \javascript{} in the browser.
More specifically, by adopting such a strict policy, we aim to quantify the
\javascript{} addiction of web elements composing a web page, through the
observation of web breakages.
As there is no standard mechanism for detecting such breakages, we introduce a
framework to inspect several page features when blocking \javascript{},
that we deploy to analyze \SI{6,384}{pages}, including landing and
internal web pages.
We discover that \percallworking{} of web pages are not strictly dependent on
\javascript{} and that more than \percmainworking{} of pages are likely to be
usable as long as the visitor only requires the content from the main section
of the page, for which the user most likely reached the page, while reducing the
number of tracking requests by~\perclesstrackingnojs{} on average.
Finally, we discuss the viability of currently browsing the web without
\javascript{} and detail multiple incentives for websites to be kept usable
without \javascript{}.
\end{abstract}

%% file: introduction.tex
\section{Introduction}\label{sec:introduction}
\javascript{} has contributed to the development of web tracking and advertisement technologies for more than two decades.
Tracking techniques have become more and more complex: simple, cookie-based tracking continues to exist, but new tracking strategies have emerged, leveraging the always-increasing complexity of the web, from browsers to server-side infrastructure.
In response, many countermeasures have been investigated and deployed, embedded into browsers or delivered as extensions.
They usually either try to prevent the browser feature from being used for tracking, for instance using site isolation~\cite{firefox-state-partitioning,chrome-cache-partitioning,DBLP:conf/www/JueckstockSSKL22}, user permission requests~\cite{firefox-rfp} or data randomization~\cite{firefox-canvas-randomization,brave-randomization}, or try to selectively block resources implementing user tracking features, while keeping the site functionality~\cite{firefox-etp}.

Blocking such tracking resources---i.e., scripts in most cases---usually relies on blocklists, which consist of rules matching requests to block.
Blocklists are a simple and efficient privacy mechanism, but are well known for suffering from several limitations.
Being maintained by humans, they tend to be limited to geographical regions and languages with enough list maintainers~\cite{DBLP:conf/www/SjostenSPPL20}.
They are also likely to be lagging behind, as trackers try to escape the blocklists~\cite{DBLP:journals/pomacs/SnyderVL20}.
There have been efforts to automate blocklist generation~\cite{DBLP:journals/popets/GugelmannHAL15}, but blocklists are also often technically limited by the type of resources they can block: aiming to selectively block the requests, they can only rely on the \acsp{url} pointing to these resources, which can be circumvented by inlining the scripts directly in the \acs{html} document or by using an always-changing, hard-to-match \acs{url}~\cite{content-blocking-livshits21}.

\javascript{} being the most powerful tracking vector, it would make it much more difficult for trackers to operate if \javascript{} was simply completely blocked in the browser.
However, \javascript{} has also many legitimate uses, and blocking it is bound to break at least some features of many websites.
Still, it is not clear whether these broken features would actually be needed by the user on most of their visits: they may be optional features which are not required at all to read or interact with the page the way the user intended to when reaching the page.

On top of this, there is no standard interface to detect whether a page element is broken, and the definition and criteria of breakage would anyway depend on its potential causes, e.g., blocking \javascript{}.

This work looks at the problem of measuring web page breakage induced by \javascript{} blocking through the following contributions:
\begin{enumerate}
  \item documenting \acs{html} code and elements that require \javascript{} to work properly,
  \item introducing a heuristic-based framework aimed at detecting potential functionality breakage introduced by blocking \javascript{}, based on limited information,
  \item performing a large-scale crawl of popular web pages, including internal pages, quantifying how badly these pages are broken when \javascript{} is blocked,
  \item semi-manually classifying page screenshots based on visual comparison,
  \item measuring the difference in request count, especially of tracking requests, when blocking \javascript{}.
\end{enumerate}

After covering some background about web tracking and advertisement relying on \javascript{} and existing tooling related to this work, we introduce our bottom-up measurement approach, detailing the page features we measure and how they can be broken without \javascript{}.
Then, we present our web crawling methodology and the results of measuring the dependency on \javascript{} of \crawlurls{}.
Finally, we detail some methodology limitations, discuss some misconceptions and incentives for websites to decrease their reliance on \javascript{} and conclude this work.

%% file: background.tex
\section{Background}\label{sec:background}
In this section, we first discuss previous work that motivates the need for reducing the amount of \javascript{} allowed client-side, then existing tooling that analyzes the usefulness of scripts sent to the user.

\subsection{\javascript{} for Advertisement and Tracking}
Previous work has shown extensively how \javascript{} is commonly used for advertisement and tracking, employing different techniques, both stateful and stateless.
In particular, \javascript{} is commonly used to handle cookies, \texttt{localStorage}, and \texttt{IndexedDB}, which do have many legitimate uses but are also notably adopted for tracking~\cite{DBLP:conf/ccs/AcarEEJND14,DBLP:conf/www/JueckstockSSKL22}, being standard, easy-to-use client-side storage technologies.

\javascript{} is also one of the core components of browser fingerprinting~\cite{DBLP:journals/tweb/LaperdrixBBA20,sensor-js-2018,laperdrix:hal-03152176, DBLP:journals/popets/RizzoTM21,mobileWebAPIAttacks2019,DBLP:journals/popets/YangY20,laperdrix:hal-01285470,DBLP:conf/sp/IqbalES21}, a technique used to collect many different browser properties and generate a distinctive fingerprint.
It can recognize the browser across sessions while storing no data on the client side, provided the fingerprint does not vary too much~\cite{DBLP:conf/imc/LiC20}.
\javascript{} also makes it easier to exploit the browser caches for fingerprinting~\cite{swNdss2021, DBLP:conf/ndss/SolomosKKP21}.
Various security attacks targeting the browser, including timing-based attacks, are able to leak data from other processes by leveraging \javascript{} for their exploitation~\cite{10.1007/978-3-319-66399-9_11}, even though it is not strictly required~\cite{DBLP:conf/uss/ShustermanAOGOY21}.

Many defenses have been deployed in popular browsers over the past few years, but there still is an ongoing arms race between browser vendors and trackers.
Even when a countermeasure is implemented, some future change can make the browser exploitable again, as highlighted by \javascript{} timing exploitation in~\cite{DBLP:conf/eurosp/RokickiML21}.

In addition, advertisement delivery heavily relies on \javascript{}~\cite{254370}, for instance to try to circumvent adblockers~\cite{DBLP:conf/ndss/ZhuHQSY18}.
One can therefore observe that \javascript{} is a known vector threatening user privacy, but that existing countermeasures keep failing at proposing a sustainable solution to preserve \javascript{} from undesirable exploitations.

\subsection{Automatic \javascript{} Removal}
To the best of our knowledge, this is the first work to detect the dependency on \javascript{} of web elements based on markup alone---without a differential analysis with the original page.
Some previous works have investigated the selective removal of \javascript{}, especially for performance purpose:
Chaqfeh~\emph{et~al.} have investigated the removal and replacement of non-essential \javascript{}, with the aim of improving performance on low-end devices~\cite{DBLP:conf/www/ChaqfehZHS20}.
The system acts as a rewriting proxy, modifying pages to remove or replace non-essential \javascript{}.
The proxy does load the whole page in the first place, including all scripts, and can thus benefit from a better understanding of the page to make informed decisions.
However, this also means that running this proxy locally would bring no privacy improvement.
Other works that classify \javascript{} usefulness based on the whole page~\cite{DBLP:journals/corr/abs-2106-13764} suffer from the same issues for this use case.

%% file: methodology.tex
\section{Page Feature Breakage}
In this section, we introduce our bottom-up analysis approach and detail the heuristics developed to measure the reliance of web page elements on \javascript{}.

\subsection{Breakage Detection}\label{subsec:breakage-detection}
To investigate the reliance of web elements on \javascript{}, we introduce a client-side measurement framework aimed at detecting potential functionality breakage when blocking \javascript{}, relying only on the \acs{dom} state after the initial page load.
This measurement framework is heavily unit-tested and is then used to measure the reliance of web page elements on \javascript{}.

\subsubsection{Limited Information Available}
We aim to be able to detect web elements present on the page that rely on \javascript{} to function.
This is ultimately useful for users willing to browse with minimal \javascript{}, to locate broken or incomplete page features, which may not always be easily spotted visually.
Thus, the detection mechanisms are restricted to inspecting the \ac{dom} state obtained after the page is fully loaded, with scripts completely blocked---i.e., the markup received from the server.
Since we must only rely on the markup of the page visited by the user, we cannot compare a possibly broken version of the page (with \javascript{} blocked) with a supposedly working one (with \javascript{} loaded), be it using \ac{dom} or visual analysis, since the scripts would then need to be downloaded and executed, defeating the privacy and security benefits.
In this setup, there is no way of knowing, from the markup alone, whether a script is meant to attach an event listener to an element (e.g., a button) to handle its possible action.

As no programming interface is available to detect breakage of web elements caused by \javascript{} blocking and based on this set of restricted information, we develop a framework of heuristics, embedded in a browser extension, that detects page features of interest and classifies them as being either working or broken.
We refrain from making any network request and from modifying the inspected page, to make it as less intrusive as possible.

Future extensions may also leverage this framework to detect and fix broken page features required by the user.

\subsubsection{Bottom-up Approach: Detecting Broken Elements to Detect Broken Features}
\acs{html} documents are built as a combination of basic \acs{html} elements.
Some of them---like \htmltag{div}---are not meant to be interactive elements, some others---like \htmltag{a}---have a native, fully functioning behavior without \javascript{}, while others---like \htmltag{canvas}---always require \javascript{} to be useful.
These basic \acs{html} elements are then combined to provide page features desired by the website developers, such as dropdown menus and accordions.
In doing so, non-interactive elements are often used or misused to provide the desired interactive, complex page features (such as using a \htmltag{div} for a button).
This makes it much harder to detect broken page features.

Adding to this impediment, there is also no general interface to detect whether basic \acs{html} elements are functionally broken, mostly because the definition and possible symptoms of breakage depend not only on the causes (e.g., blocking \javascript{}), but also on the user preferences and tolerance to partial breakage.
This means that custom heuristics, tailored to detect breakage induced by blocking \javascript{}, are required to detect broken features.
In this work, we adopt a bottom-up approach where, in place of trying to define possible symptoms of \emph{page} breakage, we instead focus on detecting breakage of web \emph{elements}, or combination of elements, when blocking \javascript{}.
Thus, breakage does not need to be defined on the whole page, only on individual web elements.

To identify possible breakage of individual web elements and of complex page features, we carried out a comprehensive analysis of standard \acs{html} elements, which can be found in \autoref{tab:js-html-elements} in the appendix, and of popular web component libraries (in \autoref{tab:js-framework-components}), aided by the \ac{waiaria} widget list~\cite{aria-widgets} and manual browsing.
We identified their respective potential reliance on \javascript{}, based on standard definitions and documentation, in-the-wild observations and common knowledge of the field regarding bad practices, and derived the custom breakage detection heuristics from this analysis.
We do not have to handle user interactions separately, as they are already part of the expected behavior of individual elements, and thus are already covered by our heuristics.

The remainder of this section details the most relevant page features and the basics of the breakage detection heuristics.

\paragraph{Images}
Standard \htmltag{img} and \htmltag{picture} elements do not require \javascript{} to work properly.
The browser renders them by directly downloading the source images from the \htmlattr{src} and \htmlattr{srcset} attributes.
We can therefore objectively define the breakage of an image element as:

\featurebox{Image}{%
Image is not rendered or a low-resolution placeholder is displayed instead.
}

Nonetheless, some websites implement image lazy loading in \javascript{}.
Instead of using the standard source attributes, they store the image source \acsp{url} in other attributes, often using custom \texttt{data-*} attributes~\cite{html-spec-custom-data-attribute,http-archive-almanac-data-attributes}, accessible with the \idlattr{dataset} \acs{idl} attribute.
Then, when the image comes close enough to the viewport, some \javascript{} logic copies the \acs{url} to the appropriate \texttt{src}/\texttt{srcset} attribute to load the image as detailed in \autoref{lst:lazy-loaded-images-js}, effectively deferring fetching the image until it is actually needed, making the initial page load faster.
In this scenario, if \javascript{} is blocked, no image will be rendered.

However, this behavior does not need to be implemented in \javascript{} anymore since the introduction of the \htmlattrvalue{lazy} value of the \htmltag{img} \htmlattr{loading} attribute in 2019--2020~\cite{mdn-img-loading-attr}, which is gradually getting adopted~\cite{http-archive-almanac-data-attributes} and implements native image lazy-loading, as shown in \autoref{lst:lazy-loaded-images-nojs}.
This native behavior of the browser is disabled when \javascript{} is disabled, to prevent tracking of the scroll position~\cite{mdn-img-loading-attr}.

Moreover, many websites implementing image lazy loading with \javascript{} use a placeholder image until the real image is loaded.
This placeholder image is often a \(1 \times 1\)~px base64-encoded \acs{gif} image supplied inline using the \texttt{data} scheme, thus requiring no extra request, while other websites use a lower-resolution image and a \acs{css} unblurring animation when the full-resolution image is loaded.

Finally, it should be noted that some websites implementing image lazy loading with \javascript{} also provide noscript fallbacks, using the \htmltag{noscript} elements, which are only interpreted when \javascript{} is disabled in the browser, allowing the image to load if \javascript{} is blocked.

We consider as \emph{large images} all images whose height and width are both greater than or equal to \largesizepx{}.

\begin{figure}
  \begin{subfigure}[t]{\linewidth}
\begin{lstlisting}[
      style=html,
      caption={Lazy loading images with \javascript{}},
      label=lst:lazy-loaded-images-js,
      language=HTML,
      ]
<img class="lazyload" data-src="image_js.jpg" alt="Image" width="300" height="90">
<script>
  const lazyImageObserver = new IntersectionObserver((entries) => {
    entries.forEach(entry => {
      if (entry.isIntersecting) {
        const lazyImage = entry.target;
        lazyImage.src = lazyImage.dataset.src;
        lazyImage.classList.remove("lazyload");
        lazyImageObserver.unobserve(lazyImage);
      }
    });
  });

  Array.from(document.querySelectorAll("img.lazyload"))
    .forEach(lazyImage => lazyImageObserver.observe(lazyImage));
</script>
\end{lstlisting}
  \end{subfigure}

  \begin{subfigure}[b]{\linewidth}
\begin{lstlisting}[
      style=html,
      caption={Lazy loading images without \javascript{}},
      label=lst:lazy-loaded-images-nojs,
      language=HTML,
      ]
<img src="image_nojs.jpg" alt="Image" loading="lazy" width="300" height="90">
\end{lstlisting}
  \end{subfigure}

  \vspace{-2ex}
  \caption{Lazy loading images}
\end{figure}

\paragraph{Forms}
Forms are one of the few interactive mechanisms able to operate without \javascript{} when implemented properly, by relying on server cooperation.

In this work, we call \emph{forms} document sections delimited by \htmltag{form} elements, containing form controls.
Forms are meant to collect data input by the user and to send them to a remote server when the form is submitted.
Various form controls that dictate the structure and logic of the form are available, most of them implemented as a \htmlattr{type} of the \htmltag{input} element, the others as separate elements, namely \htmltag{button}, \htmltag{textarea} and \htmltag{select}.

\featurebox{Form}{%
Form cannot be submitted to the server, is not submitted to the intended \acs{api} endpoint, or some form values are not submitted.
}

To submit the form, the browser builds an \acs{http} request of the type defined by the \idlattr{method} form attribute (\texttt{GET} by default), using the \acs{url} specified as the \idlattr{action} form attribute (the current page's \acs{url} by default) and the values of form controls having a \idlattr{name}, then sends this request to the server.
For the \autoref{lst:valid-form}, typing ``test'' into the search field and checking the checkbox, then clicking the Search button, will result in a \texttt{GET} request with the \acs{url} ending with the following query parameters: \texttt{/search?q=test\&check=on}.
This means that, if a form control has no \idlattr{name}, its value will not be sent; a form with at least one control having no \idlattr{name} necessarily
requires \javascript{} to handle the form, using another reference than their
\idlattr{name} to access the form controls and to either modify the page accordingly or send a request to the server.

Unfortunately, one cannot guarantee that the server will take the request into account, be it a \texttt{GET} or \texttt{POST} request.
The absence of the \htmlattr{action} attribute cannot even be used to presume of the non-handling of the form by the server, as some websites---mostly search engines---do intend to submit these requests to the form page's \acs{url}, which is the default when the \htmlattr{action} attribute is absent.

Furthermore, the form submission itself can be broken without \javascript{}.
Indeed, two separate mechanisms allow to submit a form: dedicated submission form controls and the implicit submission mechanism.
When some form controls
(\texttt{<button type="submit">}, 
\texttt{<input type="submit">}, 
\texttt{<image type="image">}) 
are part of a form, they will submit the form when activated.
However, a form can still be activated when none of these form controls are included in the form, using implicit submission.
Implicit submission allows to submit a form by hitting a key (usually ``enter'') when a text control is focused and the form has at most one single-line text control~\cite{html-spec-implicit-submission}.

Otherwise, if the form cannot be natively submitted, \javascript{} is required to either modify the page according to the form data, to trigger the form submission, or to manually send a request accordingly.

\begin{lstlisting}[
  style=html,
  language=HTML,
  caption={A valid form},
  label=lst:valid-form,
  ]
<form action="/search">
  <input type="search" name="q">
  <label>Check:
    <input type="checkbox" name="check">
  </label>
  <button>Search</button>
</form>
\end{lstlisting}

\paragraph{Lone Controls}\label{par:lone-controls}
In this work, we call \emph{lone controls} all form controls that have no form owner---i.e., that are not children of any \htmltag{form}, nor are associated to a form using their \idlattr{form} attribute.

\featurebox{Lone control}{%
Activating the control does not trigger the intended behavior.
}

Most of these lone controls require \javascript{} to be useful: to attach an event listener to them or to read their value.
The only lone controls not necessarily requiring \javascript{} are the stateful ones,
\texttt{<input type="checkbox">} 
and
\texttt{<input type="radio">}, 
because their state can be accessed from \acs{css} with the
\csspseudoclass{checked} pseudo-class, see the \textit{Disclosure Buttons}
feature and \autoref{lst:accordion-nojs} for details.

\paragraph{Empty Anchor Buttons}
An \htmltag{a} anchor button represents a hyperlink to a destination page or a section within a page.
In this work, we call \emph{empty anchor buttons} all \htmltag{a} elements that either:
\begin{itemize}
  \item have no \htmlattr{href}, \htmlattr{name} or \htmlattr{id} \acs{html} attribute\footnote{An \htmltag{a} element with no \htmlattr{href} but with a \htmlattr{name} is an obsolete, \acs{html}4 way of marking a destination for another anchor~\cite{html4-spec-anchors-links}.},
  \item have an \htmlattr{href} attribute set to the empty string,
  \item have an \htmlattr{href} attribute set to \htmlattrvalue{\#},
  \item have a \texttt{javascript:} pseudo-protocol \htmlattr{href} that is a no-operation, see \autoref{lst:empty-anchor-buttons}.
\end{itemize}

They are very often used in a discouraged way to make buttons that look like other links on the page, with the drawback that they do not convey appropriate semantics.

\featurebox{Empty anchor button}{%
Activating the anchor does not redirect to a different \acs{url}, scroll the page to the indicated part of the document or trigger the intended custom behavior.
}

This means they require \javascript{} in the same way a \htmltag{button} element does (see \textit{Lone Controls}), except when these empty anchor buttons are actually used as standard-compliant go-to-top buttons.
\acs{html}5 has indeed standardized the use of the empty \acs{url} fragment and the \texttt{top} fragment as a way to ask the browser to jump to the top of the page~\cite{html-spec-scroll-fragment-identifier}.

Some empty anchor buttons are also not used as buttons, but only for appearance consistency in a list of anchors, where this anchor has no target \acs{url}.

\begin{lstlisting}[
  style=html,
  language=HTML,
  caption={Empty anchor buttons},
  label=lst:empty-anchor-buttons,
  ]
<a href="#">Button #0</a>
<a href="javascript:void(0);">Button #1</a>
\end{lstlisting}

\paragraph{Mislinked Fragment Anchors}
Similarly, we define \emph{mislinked fragment anchors} as all \htmltag{a} elements whose
fragment is not the empty fragment but targets an element that is not found in
the page---i.e., no element has the fragment as \idlattr{id}.

\featurebox{Mislinked fragment anchor}{%
Activating the anchor does not scroll the page to the indicated part of the document or trigger the intended custom behavior.
}

Some of these elements are used as buttons, in the same way as empty anchor buttons, others are of no actual use, and would not trigger any action, even with \javascript{}.

\paragraph{Disclosure Buttons}\label{par:disclosure-buttons}
The constructs discussed above consist of a single \acs{html} element, whose intended use is standardized.
Here, we discuss more complex page features, combining several \acs{html} elements.

\begin{figure}
  \centering
  \begin{subfigure}{0.48\textwidth}
    \centering
    \includegraphics[width=\linewidth]{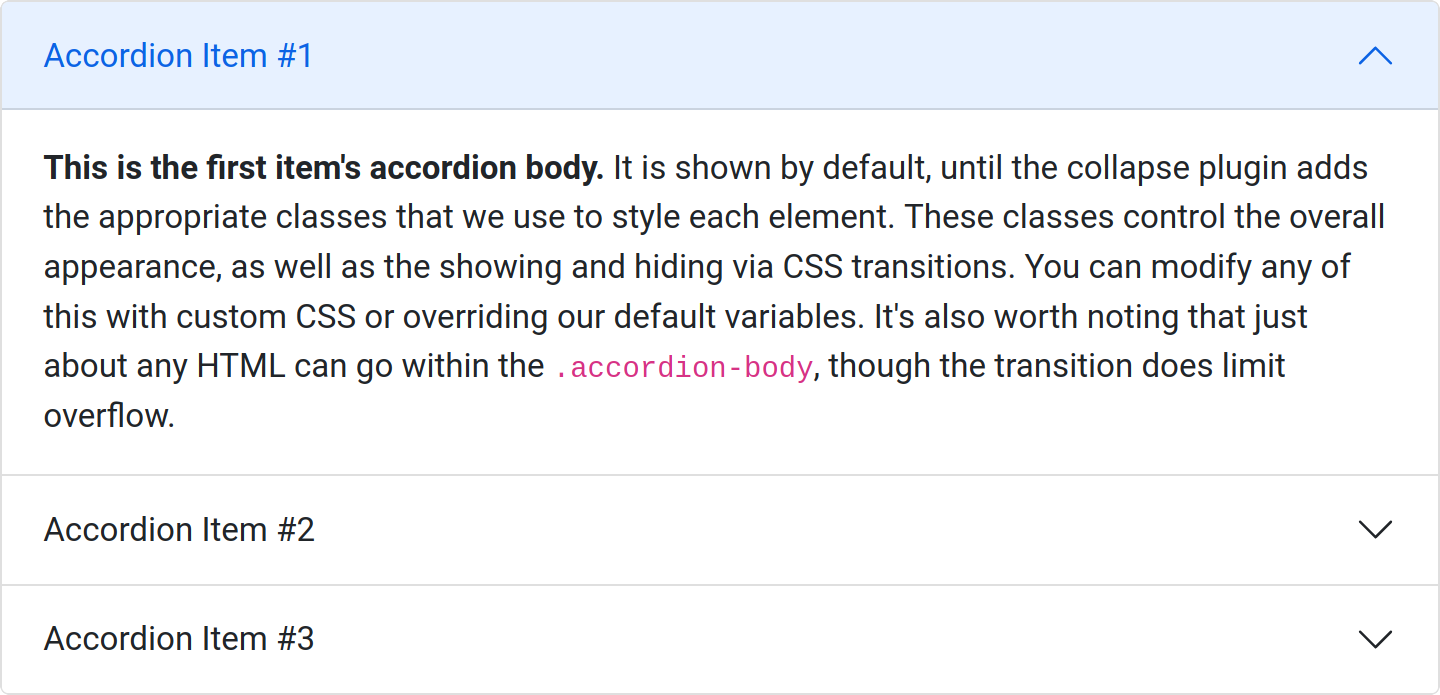}
  \end{subfigure}
  \begin{subfigure}{0.48\textwidth}
    \centering
    \includegraphics[width=0.5\linewidth]{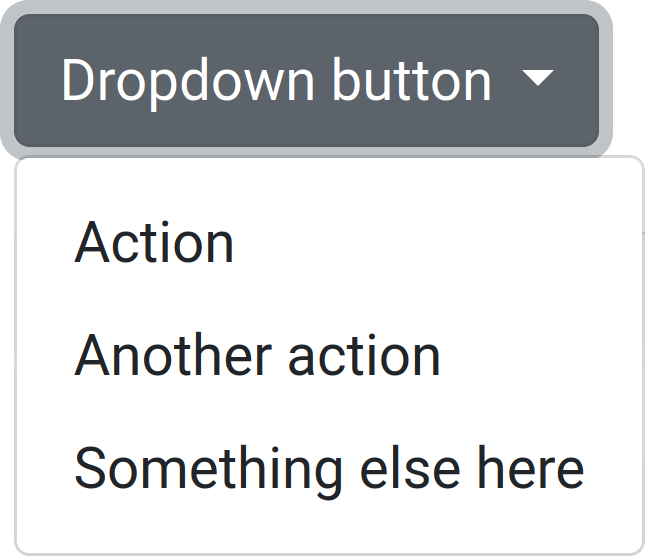}
  \end{subfigure}
  \Description{Screenshots of an accordion and an open dropdown menu}
  \caption{Examples of disclosure buttons (from the \bootstrap{}~5
  documentation~\cite{bootstrap-dropdowns,
bootstrap-accordion})}\label{lst:disclosure-button-example}
\end{figure}

We call \emph{disclosure buttons} elements having a button appearance and intended to reveal and/or hide other elements when actioned, possibly concealing information to the user if broken.
Disclosure buttons include accordion buttons and dropdown menu buttons, see \autoref{lst:disclosure-button-example}.
They are very common features, part of most popular component libraries but are also often custom components, specifically designed for the website.

\featurebox{Disclosure button}{%
Activating the disclosure button does not reveal/hide the associated element.
}

Both accordions and dropdown menus can be built without \javascript{}, but popular component libraries do require \javascript{} for these features to work~\cite{bootstrap-accordion,bootstrap-dropdowns}: an event listener is attached to the disclosure button that shows/hides the associated disclosable element when the button is activated.

A great diversity of custom implementations can be observed in the wild, using various elements for the disclosure button (usually a \htmltag{button},
\htmltag{a}, \htmltag{label} or \htmltag{div}) as well as different positions of the disclosable element relatively to the disclosure button in the
\acs{dom} tree (most of the time found as the next sibling).
Disclosable elements are most often a \htmltag{ul} (especially for dropdown menus) or a \htmltag{div}; they are sometimes dynamically inserted in \javascript{} and thus can be missing from the initial \acs{dom} tree.

Accordions and other disclosure elements can also be created without \javascript{}, using a \htmltag{label} as a button toggling an \texttt{<input type="checkbox">}, 
whose checked value can be used in a \acs{css} selector to show the disclosable element using the \csspseudoclass{checked} pseudo-class, as shown in \autoref{lst:accordion-nojs}.

\begin{lstlisting}[
  style=html,
  language=HTML,
  caption={Accordion working without \javascript{}},
  label=lst:accordion-nojs,
  ]
<div class="accordion">
  <input id="checkbox0" type=checkbox style="position:absolute;opacity:0">
  <label class="accordion-header" for="checkbox0">Header</label>
  <div id="body0" class="accordion-collapse">Body</div>
</div>
<style>
  .accordion-collapse {
    display: none;
  }
  #checkbox0:checked ~ #body0 {
    display: block;
  }
</style>
\end{lstlisting}

A simple disclosure element can also use the native element pair
\htmltag{details} and \htmltag{summary}, see \autoref{lst:details-summary}: elements in the \htmltag{details} are hidden by default, while the \htmltag{summary} element is shown and adjoined by an arrow suggesting some content is hidden; when clicking the \htmltag{summary} element, the hidden content visibility is toggled.

\begin{lstlisting}[
  style=html,
  language=HTML,
  caption={Native disclosure element},
  label=lst:details-summary,
  ]
<details>
  <summary>Some question</summary>
  Some detailed answer that is hidden by default
  and toggled by clicking the summary element.
</details>
\end{lstlisting}

The diversity of implementations and the use of non-semantic elements make it a challenge to reliably detect all disclosure buttons while minimizing the number of false positives.
Using the \acs{css} class names can sometimes help to refine the classification, especially when these are from the most popular component libraries (like \cssclass{dropdown-toggle} and \cssclass{dropdown-menu}), but this is no silver bullet, some websites using class names in the website's language, while others obfuscate the class names.

\paragraph{Protected E-Mails}
\mbox{}

\featurebox{Protected e-mail}{%
E-mail addresses (and sometimes, other strings with an at sign) are replaced by \emailprotected{}.
}

Some websites try to prevent mass harvesting of e-mail addresses by requiring \javascript{}.
They embed the encoded address, often visually replaced by \emailprotected{}, which is then decoded and displayed in place of this message.
This is an example of a feature that deliberately relies on \javascript{}.

\paragraph{Loader Overlays}
Some pages display an overlay that covers the entirety of the page until the page is loaded.
We call them \emph{loader overlays} (they are also referenced as \acs{ajax} loaders or preloaders).

\featurebox{Loader overlay}{%
The actual page content is hidden behind an overlay, which usually features a loading spinner.
}

Besides going against the flow of best practices, especially progressive loading, these overlay elements are only removed with \javascript{} when the page is done loading, meaning that, when \javascript{} is disabled, the overlay is never hidden and makes it impossible to read the page content, actually often properly loaded, even without \javascript{}.

Loader overlays usually appear as a \htmltag{div} and as a direct child of the \htmltag{body} element (often being its first child) and have the \htmlattr{id} \htmlattrvalue{preloader} or a \htmlattr{class} containing the word \htmlattrvalue{preloader}.

\paragraph{Page Text}
\mbox{}

\featurebox{Page text}{%
The page has no text content.
}

This heuristic checks whether the \htmltag{body} has some text content (using the \idlattr{textContent} and \idlattr{innerHTML} properties), other than
whitespace.

This is particularly useful when the page is a full-page app.
Full-page apps are websites where the whole page is nested in a single element that is completely populated in \javascript{}.
Without \javascript{}, the page is left blank and broken.

This heuristic applies to the whole page and is an all-or-nothing heuristic.

\paragraph{Stylesheets Loaded}
\mbox{}

\featurebox{Stylesheets loaded}{%
The page has no style loaded.
}

Finally, this heuristic detects if the page has at least some basic stylesheet loaded by checking font styles of a few elements, especially
headers, which are almost always changed by websites.

This heuristic applies to the whole page and is an all-or-nothing heuristic as well.

\subsection{Page Feature Relevance}
\begin{figure}
  \includegraphics[width=0.7in]{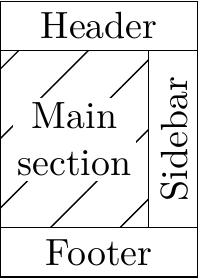}
  \Description{Diagram of the common page structure, with a header and footer, and a sidebar along the main section}
  \caption{Common page structure (the sidebar can be on either side of the
  page)}\label{fig:common-page-structure}
\end{figure}

A web page is composed of different sections with each their own purposes, as can be seen in \autoref{fig:common-page-structure}.
The header is often used to guide the navigation on a site whereas the main section offers the bulk of the site content.
In order to provide a more focused analysis of page breakage, we look to identify these sections in the pages that we crawled.

These sections can be inferred from the \acs{dom} tree, using tag names, element \idlattr{id}s, and class names.
In particular, \acs{html}5 provides semantic elements for these sections: \htmltag{header}, \htmltag{footer}, \htmltag{aside}, and \htmltag{main}.
Not all websites use these semantic elements, but many do mark these sections with a combination of non-obfuscated \idlattr{id}s and class names, which makes it possible to classify page elements according to their position and intended purpose.

In addition, some page features may never be accepted by the user, especially tracking and advertisement features, which do require \javascript{} most of the time.

\section{Data Collection}
This section details the crawling methodology we implemented to measure the dependency of web page elements on \javascript{} at scale.

\subsection{Crawled Websites and Pages}\label{subsubsec:url-choice}
We used a subset of the \hispar{} list~\cite{DBLP:conf/imc/Aqeel0FM20}, which provides a list of popular webpage \acs{url}s, including landing and internal ones.
The \hispar{} list is built using the \alexatoponem{} domain list, querying the paid \acs{api} of \googlesearch{} with the query \texttt{site:\(\omega\)} for each domain \(\omega\) and storing the first 50~search results along with their ranking, stopping as soon as the final list has \SI{100,000}{\acs{url}s}.
The user's location is set to the United States and results are limited to pages in English.
We used the most recent list that was released at the end of January 2021~\cite{hispar-list}, and focused on the first three \acs{url}s (based on \googlesearch{} rankings) of each domain, which usually include the landing page and two internal pages.
We limit the crawl to two internal pages because pages of lower ranks are very often of the same type---e.g., product pages of a shopping website.
This makes up for \crawlurls{} to crawl, from \crawldomains{} (\crawlalexabasedomains{}).
Visiting internal pages is very important in this study, since page features may be drastically different between the landing page and internal ones---e.g., the website could feature a carousel on the landing page and forms in the internal pages.

We implemented the breakage detection heuristics detailed in \autoref{subsec:breakage-detection} as a \javascript{} library intended to be used as part of a browser \webextensions{} extension.
This library detects page elements matching features of interest and classifies them as either being working or broken.

\begin{figure}
  \centering
  \includegraphics[width=0.9\linewidth]{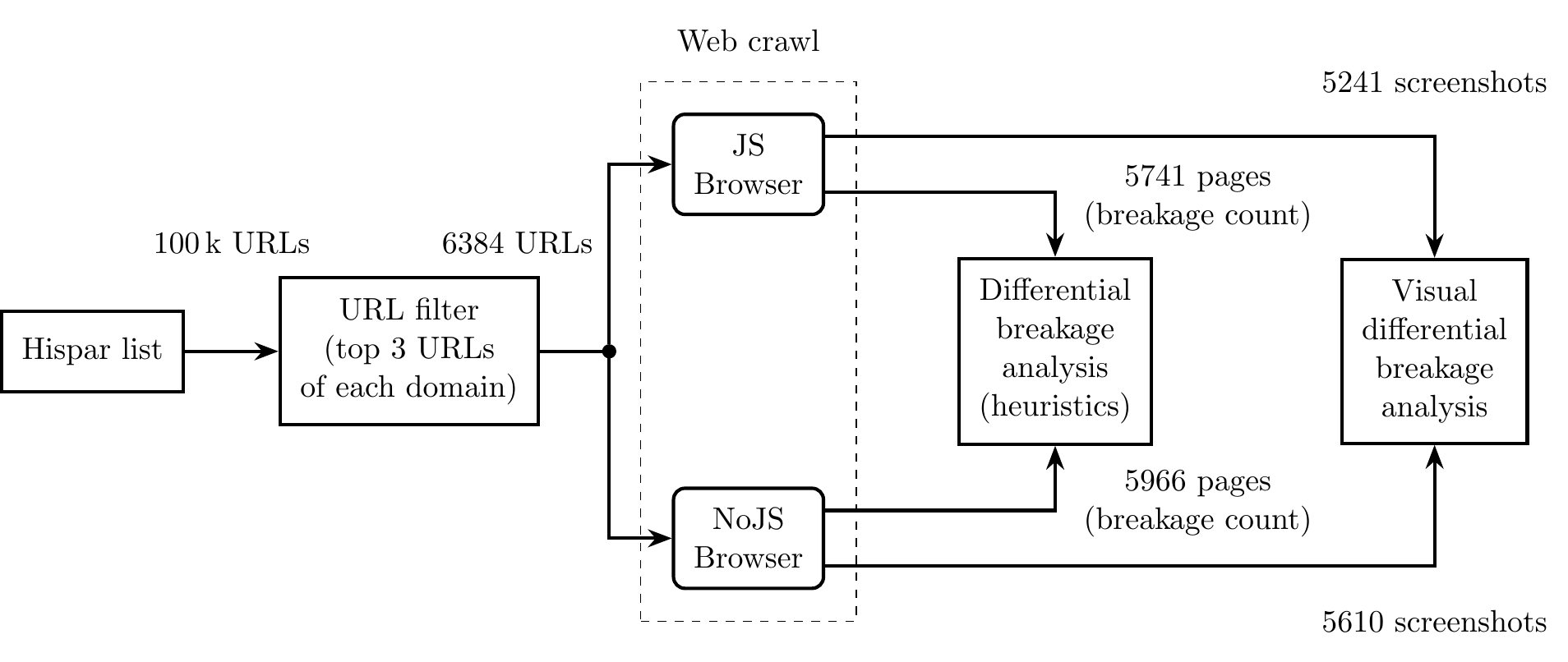}
  \Description{Diagram of the dataflow. Domains from the Hispar list are filtered then supplied to the js and nojs browsers.
  The heuristics results and screenshots are then compared between these two versions.}
  \caption{Dataflow diagram}\label{fig:dataflow-diagram}
\end{figure}

Web crawling is then automated with \puppeteer{}, using \crawlbrowser{}, required for \puppeteer{} compatibility.
The crawl uses two instances of \firefox{} running at the same time, with the breakage-detection extension loaded: one is using default preferences while the other has \firefoxpref{javascript.enabled} set to \texttt{false}, which disables \javascript{} globally for this instance, as depicted in \autoref{fig:dataflow-diagram}.
For each page \acs{url} to crawl, the page is requested in a new tab in each browser instance at the same time and the following steps are followed:
\begin{enumerate}
  \item load the web page (with a \loadtimeout{} timeout) then wait for \idletime{},
  \item inspect the loaded web page for feature breakage,
  \item save a page screenshot and the current \acs{dom} state as \acs{html}.
\end{enumerate}
The browser tab is then closed and the crawl moves to the next page \acs{url} as soon as both instances are done with the current one.
We have checked, with the help of the generated screenshots, that \idletime{} were enough for lazy-loaded content to be loaded.
The crawl was run from our campus network, from \crawlstartdate{} to \crawlenddate{}.
If the \event{DOMContentLoaded} event (indicating the initial \acs{html} document has been loaded) is not fired after~\domloadtimeout{}, the \acs{url} is skipped for this instance; if the \event{load} event is not fired after~\loadtimeout{}, the \acs{url} is skipped as well.
The crawl waits for~\idletime{} to leave enough time for asynchronous content to load, especially stylesheets and lazy-loaded images.
Finally, the extension is given \inspectiontimeout{} for the breakage-detection inspection.

The crawl with default preferences is hereafter referred to as \crawlplain{}, while the one with \javascript{} disabled as \crawlnojs{}.
For both these crawls and for each page, the following data are collected:
\begin{itemize}
  \item \emph{counts of broken and working elements} for each page feature, in the main section and in the whole page, and whether these elements are visible or not, forming a \acs{json} feature report,
  \item \emph{a page screenshot} for further, manual analysis,
  \item \emph{the page \acs{dom}}, as an \acs{html} file.
\end{itemize}

\subsection{Collection of Website Categories}
To discover and highlight breakage disparities between website categories, we adopted the classification from \opendns{}~\cite{opendns-domaintagging}, which uses a crowdsourcing system to attribute categories to domains.
Registered users can propose domains and vote on categories.
To the best of our knowledge, it is the only website classification service that provides definitions of categories (as this is required for crowdsourcing, definitions are available at~\cite{opendns-domaintagging-categories}) and that can attribute several categories to each website---i.e., categories are not disjoint.
This comes with the drawback that only about \percdomainscat{} of domains from the dataset are covered.
Some highly correlated categories are merged into new ones to improve readability.

%% file: results.tex
\section{Results}
In this section, we report the results of crawling the subset of the \hispar{} list built above: \crawlurls{} from \crawldomains{} (\crawlalexabasedomains{}).

\subsection{Dataset Description}
\autoref{tab:crawl-stats} presents the data collected during the crawl, and the result of each crawl step.
Out of the \crawlurls{} crawled, \crawlurlsbothloaded{} were successfully loaded, feature breakage data was collected for \crawlurlsbothfeatures{} and the \acs{html} was saved for \crawlurlsbothhtml{} for the same crawl \acs{url} for both \texttt{[plain]} and \texttt{[nojs]}
crawls.

\begin{table}[H]
  \centering
  \caption{Success statistics for each crawl step}\label{tab:crawl-stats}
  \input{page_stats.tex}
\end{table}

Inability to load and process some of the pages can be attributed to various factors, including region-based blocking~\cite{DBLP:conf/uss/TschantzASQJP18}, possibly outdated \acsp{url} of the \hispar{} list at the time of crawl or load and processing timeouts due to long synchronous \javascript{} rendering times and heavy page \javascript{} processing.

\subsection{Effect on Page Load Time}\label{subsec:page-load-time}
\begin{figure}
  \includegraphics[width=0.5\linewidth]{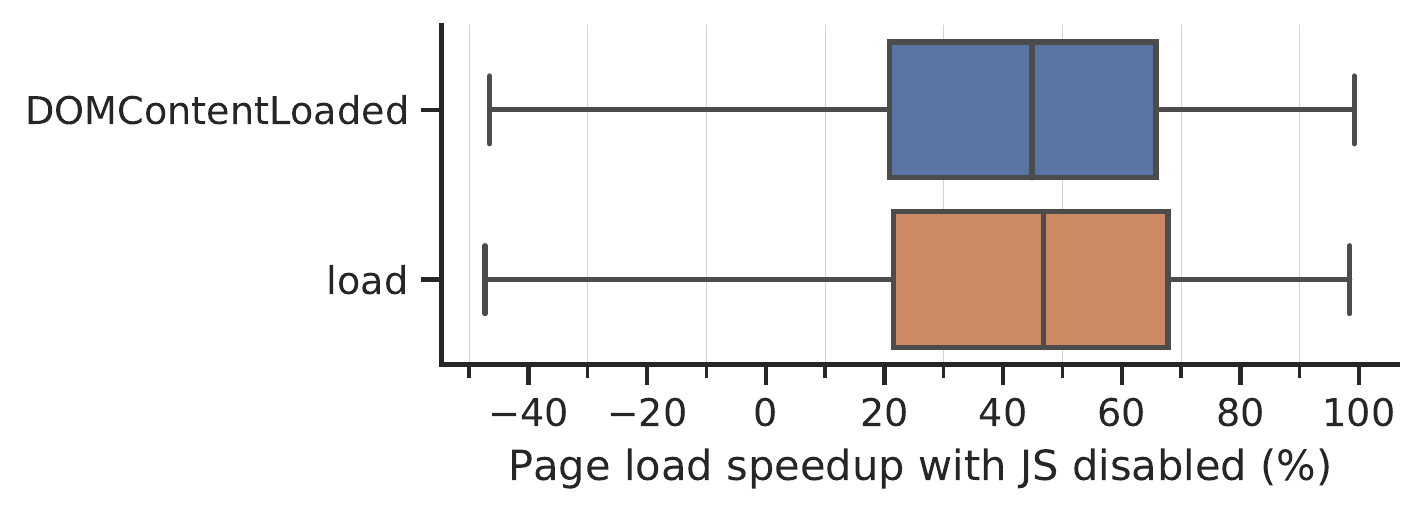}
  \Description{Boxplots of speedup when javascript is disabled: the median speedup is more than 45\%}
  \caption{Speedup when blocking \javascript{}}\label{fig:nojs-speedup}
\end{figure}

Blocking \javascript{} brings up a significant page load speedup for most pages, as shown in \autoref{fig:nojs-speedup}.
In our dataset, the median load time with \javascript{} enabled is \medianloadtimeplain{}, while it is reduced to \medianloadtimenojs{} when disabling \javascript{}.
This can be attributed to \javascript{} files not being downloaded, parsed and executed, and to some content not being loaded by \javascript{}, especially lazy-loaded images that do not provide a fallback.

Some pages are however slower to load, because of the event we used to detect the page load completion.
The \event{load} event is fired by the browser as soon as the whole page is loaded, but this does not account for content downloaded in \javascript{}, such as lazy-loaded images or font files.
Moreover, as image lazy loading is not possible when \javascript{} is disabled in the browser (the \texttt{loading="lazy"} 
standard behavior is disabled by the browser to prevent tracking), image load time is then included in this \event{load} event, explaining the higher load times measured when \javascript{} is disabled.

\subsection{Page Section Classification}
In this section, we validate the possibility of discovering the different sections of a page from the standard, semantic metadata built in the \acs{html} document.
Notably, we look for either explicit \acs{html} elements, \htmlattr{id}s, or classes that relate to the structure of a page.

\autoref{tab:page-section-stats} details the distribution of pages from the dataset for which we were able to identify at least one of the specified selectors (\textit{element}, \textit{\#id}, \textit{.class}).
The ratio of pages having a \htmltag{main} element matches 2021 \httparchive{}'s findings very well~\cite{http-archive-almanac-main-tag-2021}.
Around \perckindamainsec{} of pages have clearly identifiable main sections, while more than \perckindasemantic{} of pages mark the main section, header or footer of the page, making it possible to recover the remaining main section.
In the following, when the main section cannot be determined, the whole page is considered as being the main section.

\begin{table*}
  \footnotesize
  \centering
  \caption{Ratios of pages where the following \acs{css} selectors match at
  least one element}\label{tab:page-section-stats}
  \input{page_section_stats_table.tex}
\end{table*}

\subsection{Page Feature Breakage}
Then, we report about page dependency on \javascript{} and we detail observed feature breakage when disabling \javascript{}.

We start by introducing the quantities used to quantify feature breakage when blocking \javascript{}.
First, the differential breakage (\acs{dbr}) of a page feature is the difference between the counts of broken elements with \javascript{} disabled and with \javascript{} enabled~:
\begin{equation}
  \dbr{feat}{page} =
    \brokencount{\crawlnojs{}}{feat}{page}
    - \brokencount{\crawlplain{}}{feat}{page}
\end{equation}

A high, positive \acs{dbr} denotes that the page has many more broken elements (of the relevant feature) when blocking \javascript{} than with
\javascript{} enabled.
The differential breakage can become negative when the \crawlnojs{} page has less elements detected as broken than the \crawlplain{} page.
This may happen in different scenarios, including when:
\begin{itemize}
  \item elements that are considered as broken are dynamically added in \javascript{} (e.g., in single-page applications) and are thus not present  in the \crawlnojs{} page,
  \item elements are hidden and only become visible with \javascript{} (if the differential breakage is restricted to visible elements only),
  \item the page provides noscript fallbacks which are detected as working while
    the \crawlplain{} page has elements that cannot be detected as working or are actually broken.
\end{itemize}

Then, since elements are labeled as either broken or working, the total count of a feature is the sum of their counts:
\begin{equation}
  \totcnt{\crawlnojs{}}{feat}{page} = \brokencount{\crawlnojs{}}{feat}{page} + \workingcount{\crawlnojs{}}{feat}{page}
\end{equation}

Finally, the normalized differential breakage (\acs{dbrn}) can be derived from these two quantities:
\begin{equation}
  \dbrn{feat}{page} =
  \begin{cases}
    \frac{\dbr{feat}{page}}{\totcnt{\crawlnojs{}}{feat}{page}}
      & \text{if } \totcnt{\crawlnojs{}}{feat}{page} \neq 0\\
    0 & \text{otherwise}
  \end{cases}
\end{equation}

\subsubsection{Aggregated Features}
To ease the interpretation of elementary page features, we define two aggregate features as the following unions:

Interactive features = \{
Lone Controls,
Forms,
Empty Anchor Buttons,
Mislinked Fragment Anchors,
Disclosure Buttons~\}

and

Main features = \{
Page Text,
Stylesheets Loaded,
Interactive features,
Large Images,
Loader Overlays~\}.

As not all the features included in the \textit{Main features} have the same weight (the fact that no large image is broken does not matter if the page has no text content), the maximum of each metric is taken when aggregating, so that the \textit{Main features} feature gives the more reasonable classification of whether the page is broken or not.

\subsubsection{Disparity Across Website Categories}
\begin{figure*}
  \includegraphics[width=\linewidth]{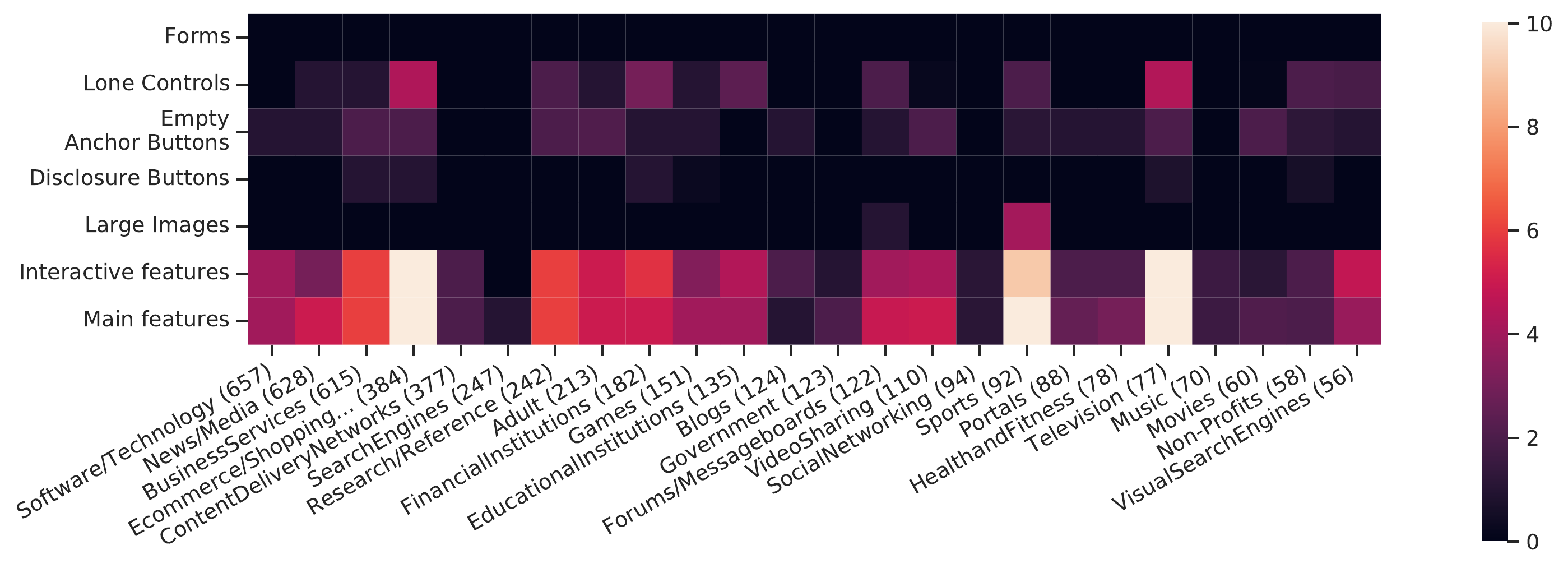}
  \Description{Heatmap of differential breakage of visible elements in the main section, across website categories. Blogs are much less broken than e-commerce sites when blocking javascript.}
  \caption{Color represents the 90\textsuperscript{th} percentile of differential breakage of visible elements in the \textit{main section}.
  Lighter shades denote higher differential breakage.
  Only categories with more than \catthreshold{} in the dataset are plotted, to improve readability. This highlights the disparity of breakage across website categories.}\label{fig:features-main-dist}
\end{figure*}

\autoref{fig:features-main-dist} reports on the 90\textsuperscript{th}~percentile of differential breakage of visible elements \textit{from the main section} for each feature and across page website categories.
This figure highlights the disparity of breakage observed in the main section across website categories.
In particular, e-shopping pages have more broken interactive features in the main section than blogs have, mostly because they contain more interactive elements in this section (\SI{25.5}{\text{interactive elements}} for \texttt{[plain]} e-shopping pages and \SI{8.5}{\text{interactive elements}} for \texttt{plain} blogs on average, in the main section): e-shopping pages can have an order form and other interactive elements (to build light boxes for example) on a product page, for instance.
\autoref{fig:features-all-dist} from the appendix takes into account the whole page for comparison.

\subsubsection{Dependency of Main Page Features on \javascript{}}
\paragraph{Reliance on \javascript{}}
Every crawled page has at least one \htmltag{script} with \idlattr{type} \texttt{text/javascript}.
The page with the highest count of these tags has almost 300~of them.

\begin{figure}
  \includegraphics[width=0.7\linewidth]{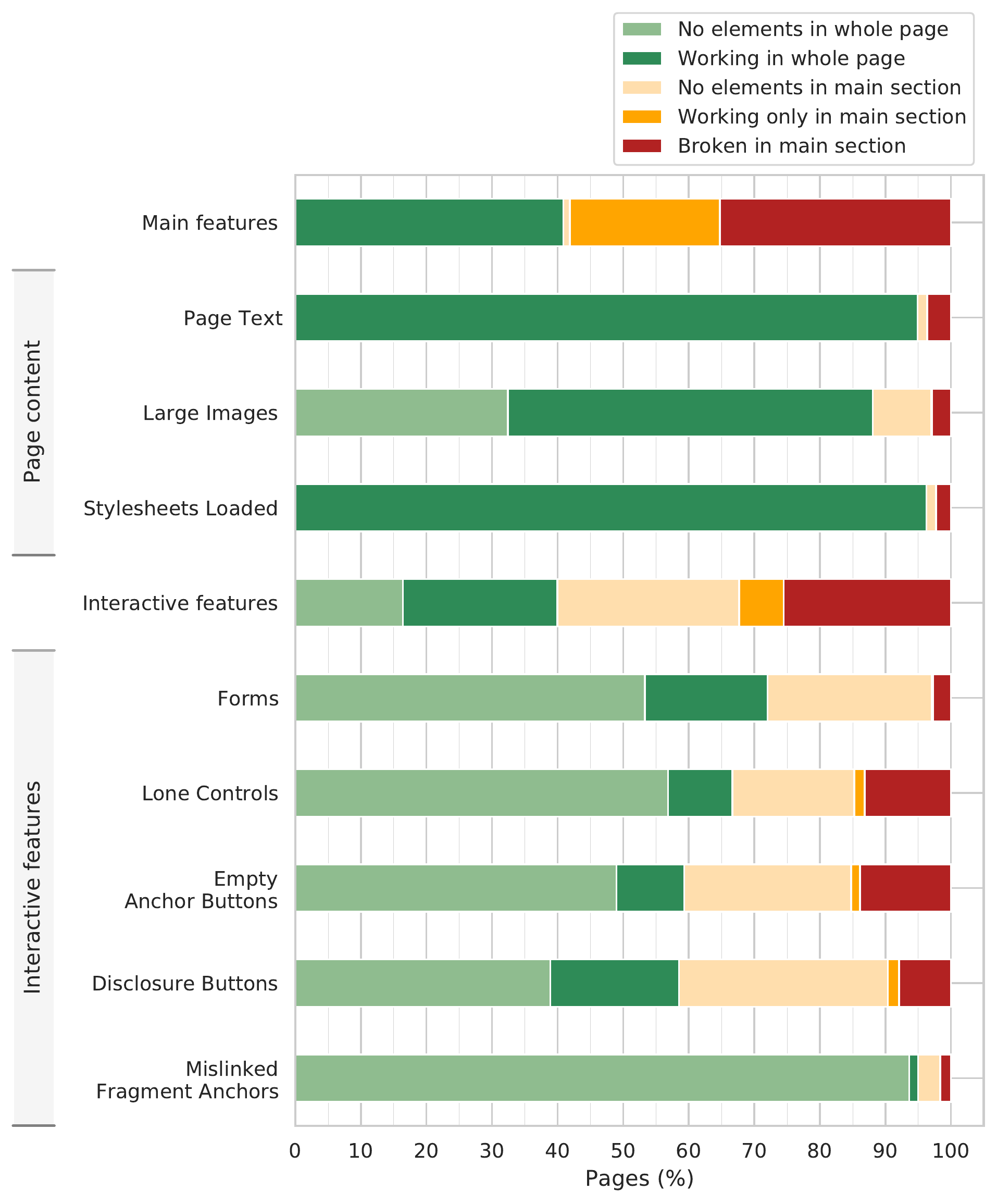}
  \Description{Barplot of page shares showing whether page features are broken or not.}
  \caption{Shares of pages for each page feature status; when present, a feature can either be working in the whole page, only in the main section or broken in the main section}\label{fig:bar-broken-status}
\end{figure}

\paragraph{Feature Breakage Report}
\autoref{fig:bar-broken-status} reports on the proportions of pages for each page feature possible status, indicating if these features are broken or working, or if there are no elements matching this feature (which is thus not broken).
It can be seen that \percmainworking{} of pages have all their main features from the main section working, which means they are likely to be useful to the user, even with \javascript{} disabled.
Even when taking the whole page into account, \percallworking{} of pages have all their main features still working, meaning that the whole page is likely working as intended, when blocking \javascript{}.
This difference between the main section and the whole page is easily explained by the fact that, on many pages, interactive features are used around the main section---e.g., for navigation--- not in the actual content, as quantified in \autoref{fig:bar-broken-status}.
However, it should also be noted that for around \percintermain{} of pages, at least one element from an interactive feature found in the main section is broken, which could impede the user from using the page as intended.

\begin{figure}
  \includegraphics[width=0.7\linewidth]{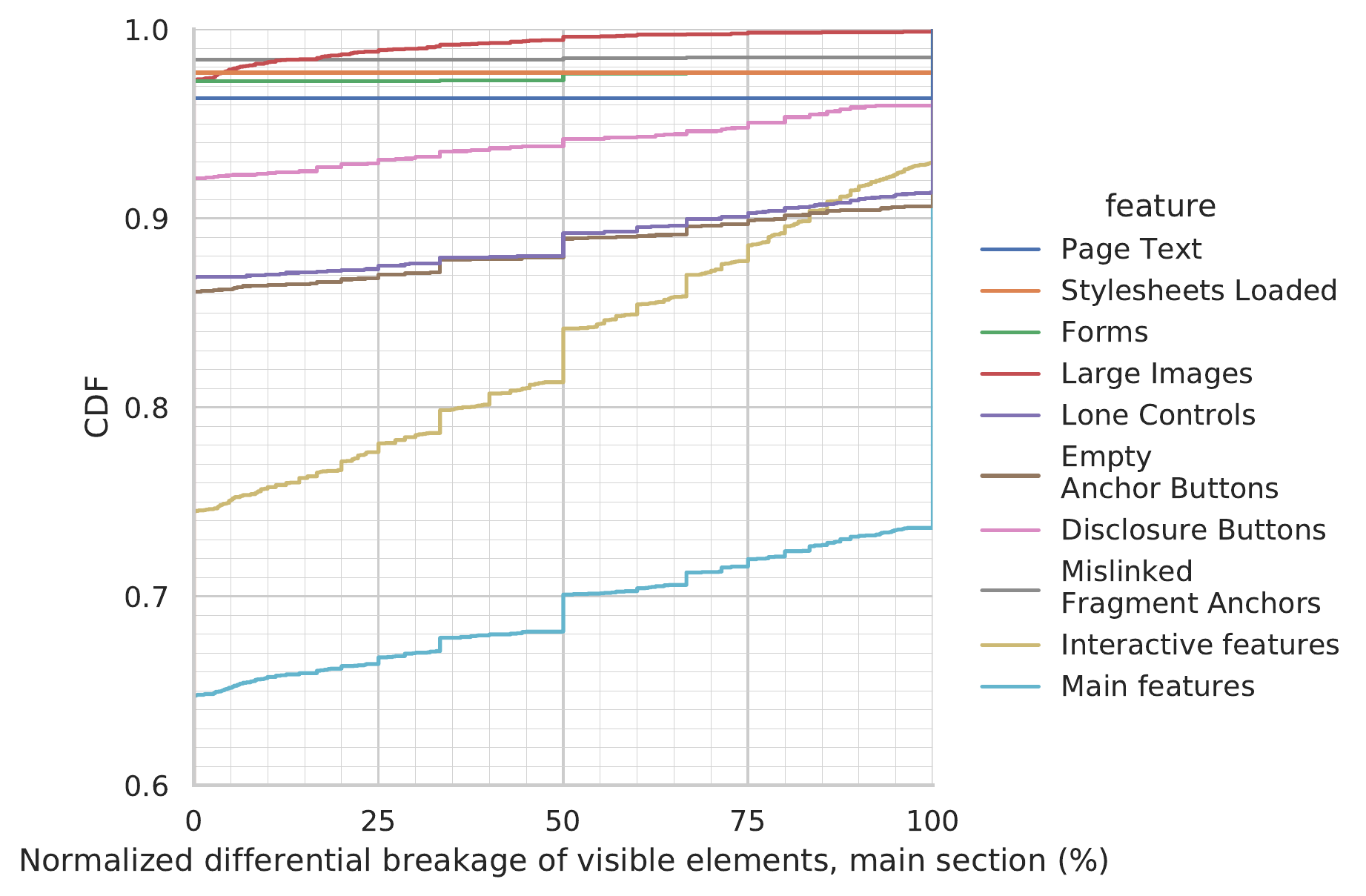}
  \Description{Line plot of the normalized differential breakage of visible elements of the main section. It highlights that when some elements of a page feature are broken, most of them usually are.}
  \caption{Normalized differential breakage (\acs{dbrn}) of visible elements of the \textit{main section}; negative normalized differential breakage is not shown for readability}\label{fig:main-perc-dist}
\end{figure}

\paragraph{Feature Implementation Consistency}
\autoref{fig:main-perc-dist} depicts the normalized differential breakage (\acs{dbrn}) of visible elements of the main section, a high \acs{dbrn} meaning that most of the elements matching a feature are broken.
\autoref{fig:main-perc-dist} highlights that, when an elementary interactive feature is at least partially broken on a page, it is actually completely broken most of the time (short transitions between \SI{0}{\%} and \SI{100}{\%}~\acs{dbrn})---i.e., all elements of this feature are broken.
Beyond the fact that the main section usually contains few interactive elements, this can be attributed to the fact that the implementation of the different elements of a feature on a given page is likely to be the same, especially when the page is using a \acs{ui} framework (e.g., \bootstrap{}~\cite{bootstrap-homepage}), that standardizes the implementation.
In other words, for a given feature, it is very unlikely that a given page has elements that require \javascript{} while others do not.

%% file: page_stats.tex
\begin{tabular}{llll}
\toprule
Crawl type &             [plain] &              [nojs] &                both \\
Page           &                     &                     &                     \\
\midrule
Crawled        &                6384 &                6384 &                6384 \\
Loaded (1)     &  5875 (\SI{92}{\%}) &  6119 (\SI{96}{\%}) &  5827 (\SI{91}{\%}) \\
Inspected (2)  &  5741 (\SI{90}{\%}) &  5966 (\SI{93}{\%}) &  5695 (\SI{89}{\%}) \\
HTML saved (3) &  5241 (\SI{82}{\%}) &  5610 (\SI{88}{\%}) &  4911 (\SI{77}{\%}) \\
\bottomrule
\end{tabular}

%% file: page_section_stats_table.tex
\begin{tabular}{lrr}
\toprule
                                                                           Selector &  Page count &  Ratio (\%) \\
\midrule
                                                                      \texttt{main} &              1557 &        27.6 \\
                                                       \texttt{main, \#main, .main} &              2128 &        37.8 \\
                                                      \texttt{main, header, footer} &              3653 &        64.8 \\
                        \texttt{main, header, footer, aside, nav, section, article} &              4097 &        72.7 \\
 \texttt{main, \#main, .main, header, \#header, .header, footer, \#footer, .footer} &              4311 &        76.5 \\
\bottomrule
\end{tabular}

%% file: limitations.tex
\section{Limitations}
In this section, we detail the limitations of the heuristic framework and of our crawling methodology.

\subsection{Measurement Framework Limitations}
\subsubsection{Limited Information Available}
Since our heuristic-based measurement is intended to only rely on information available when no \javascript{} is loaded, it is limited to what is discoverable based on the initial state of the \acs{dom} only.
It is thus unable to detect the reliance on \javascript{} of elements whose such reliance cannot be derived form the markup, mostly some cases of misuse of non-semantic elements, e.g., \htmltag{div}s used to build buttons.

\subsubsection{User Expected Feature Granularity}
The heuristics that comprise the measurement framework are derived from basic \acs{html} elements and common components, following a bottom-up approach to detect the reliance of web elements on \javascript{}.
In some cases, this might not match the level of granularity of features expected by users, the framework being in those cases more fine-grained---e.g., only buttons of a modal would be reported broken, not the modal itself.
This limitation results from the limited information available and from the web platform, since the relationship between elements is usually not automatically discoverable based on static analysis---e.g., it is usually impossible to discover that a button is used to close a modal, based on markup alone.

\subsection{Crawl Limitations}
The measurement crawl is limited to three \acsp{url} per domain, which does not cover the whole site, but we believe this still provides reasonable insight about reliance on \javascript{}, especially because many webpages from a website actually follow the exact same template.
In addition, the crawl is run from a single location, from a single device, on desktop, but we expect few differences on mobile since the usage of \javascript{} is similar~\cite{http-archive-almanac-js-usage-2021}, except for components only shown on mobile, such as hamburger menus.
We also expect very few differences between browsers and devices, since we browse without \javascript{}, the interface of the web platform being very similar when \javascript{} is not enabled.

%% file: discussion.tex
\section{Discussion}
In this section, we discuss some misconceptions about current reliance on \javascript{}, the viability of \javascript{} blocking, and incentives for website owners for making their websites usable without \javascript{}.

\subsection{Manual Visual Analysis of Screenshots}
To back up these results, which can seem to go against the common assumption
that web pages are utterly broken without \javascript{}, we manually labeled
page full-page screenshots collected during the crawls, according to their
breakage seriousness.

To this end, we first automatically compared all \crawlplain{} and \crawlnojs{} page
full-page screenshots pixel-to-pixel, automatically detecting and disregarding most of the differences due to
vertical shift of content between the two versions.
Screenshots were then put into \SI{10}{\%}-wide bins according to their pixel-to-pixel
difference percentage.
As pixel-to-pixel difference would greatly overestimate the actual,
user-perceived page difference, and, in some cases, underestimate the user-perceived difference on pages where the background takes the most room, one of the author manually labeled \SI{150}{random~samples} in each of the four bins with the smallest difference,
according to the amount of information lost to the user when disabling
\javascript{}, by visual comparison only.
Missing advertisements, cookie banners or presentational-only images (not
bringing specific information) are not considered information loss, while
missing information-heavy images or substantial layout breakage are.

Based on the labeling we conducted and using the pixel-to-pixel difference bins previously constructed, we concluded
that at least \SI{50}{\%} of pages feature no substantial amount of information loss---such as missing content text or figure that could mislead the user---when blocking \javascript{}.
This lower bound is negatively impacted by page differences resulting in a
significant pixel-to-pixel difference percentage while actually making the page
more usable when blocking \javascript{}, a major example being semi-transparent
cookie-consent overlays that cover the entire viewport, which do not appear when
blocking \javascript{}.

\subsection{\javascript{} Reliance of Most Visited Websites and Component Framework Trends}
The common assumption that web pages heavily depend on \javascript{} may stem
from the fact that most visited websites do heavily rely on \javascript{}.
For instance, \youtube{} and other \google{} products, \twitter{} and
\instagram{} are largely unusable when \javascript{} is blocked: e.g., the
\youtube{} landing page displays only loading placeholders in that case.

Moreover, deployment of relatively recent \javascript{} client-side component
frameworks such as \react{} or \vuejs{}, which may result in a blank page
without \javascript{}, is misrepresented by developer trends.
These projects are indeed among the top~10 repositories on \github{} based on
star rankings~\cite{github-star-rankings-repo}, but constitute only a very small
share of websites actually deployed, \react{}
and \vuejs{} accounting respectively for less than \SI{3}{\%} and \SI{1}{\%} of
websites monitored by \wcubetechs{}~\cite{w3techs-js-libs-usage-stats}.

\subsection{Benefits and Viability of Aggressive \javascript{} Blocking}
On top of the numerous privacy and security benefits introduced in \autoref{sec:introduction} and \autoref{sec:background}, disabling \javascript{} brings additional changes of different natures.

\subsubsection{Tracking Reduction Benefits}
To understand the impact of disabling \javascript{} on a user's online footprint, we performed a new crawl on the same set of \acsp{url} where we logged the number of requests triggered by each web page.
We leverage the \texttt{onBeforeRequest} handler~\cite{webext-onbeforerequest} of \firefox{} to classify each \acs{url} in real-time.
The classification relies on \firefoxetp{} and the built-in \disconnect{} lists~\cite{firefox-etp} to indicate if a \acs{url} is either first or third party and whether it is involved in tracking or not.

\autoref{tab:mean-request-count} details the result of this new crawl.
Considering all types of requests, blocking \javascript{} presents a mean reduction of \SI{61}{\%} and this percentage is even higher at \SI{85}{\%} for tracking requests.
This shows how beneficial disabling \javascript{} can be when it comes to tracking.
Looking at the difference between first and third party requests, we can notice a difference in the type of loaded resources.
While images and stylesheets are mostly loaded in a first party context, scripts mostly come from third parties and blocking \javascript{} here has a drastic impact, as these are never loaded in the browser.
Some \acsp{xhr} are also preloaded using \texttt{<link rel="preload" as="fetch" src="\dots">}, 
which explains why a few \acsp{xhr} are still detected with \javascript{} disabled.

\begin{table}
  \centering
  \caption{Mean request count and standard deviation for each request type with \javascript{} enabled and disabled}\label{tab:mean-request-count}
  \input{crawl_requests_table.tex}
\end{table}

\subsubsection{Browsing Comfort}
In the case where the page is usable enough without \javascript{}, having it disabled can actually improve the browsing experience by reducing the amount of aggressive page behaviors, such as pop-up advertisements, newsletter forms or unexpected animations, resulting in less obstructed browsing.
In particular, cookie banners, that ask for user consent, are usually not shown in this scenario since they are often completely managed with \javascript{} cookie consent frameworks, that set the cookies in \javascript{}.

\subsubsection{Reduced Data Size}
Since external scripts are not loaded when \javascript{} is disabled, this can result in a sizable reduction of transferred data; the \httparchive{} reporting a median size of \SI{444}{kB} of \javascript{} per page~\cite{http-archive-almanac-javascript-usage}.

Reducing the amount of transferred data has several benefits, including faster page loads for most pages (see \autoref{subsec:page-load-time}), reduced
connection-related energy consumption, especially on mobile devices, and reduced cost for users with a data cap on their mobile plan.
However, in some cases, disabling \javascript{} could actually result in higher bandwidth usage, as it would prevent loading some pieces of content only when actually needed, as with lazy-loaded images.
For instance, since all images would be loaded disregarding of the scroll position, it could happen that the user would actually leave the page without ever reaching its bottom where images would have been needlessly downloaded.

\subsubsection{Reduced Client-Side Processing Load}
Reducing the amount of scripts processed on the client side also reduces the processing stress on the end device.
This results in a lower consumption which, especially on mobile devices, can increase battery life and device life span due to reduced heating and battery stress, reducing e-waste.
Varvello and Livshits have tested 15 Android browsers and found that ad blocking can offer between \SI{20}{\%} to \SI{40}{\%} of battery savings with an additional \SI{10}{\%} when dark mode is enabled~\cite{DBLP:journals/corr/abs-2009-03740}.

\subsubsection{Impediments to Non-\javascript{} Browsing}
Despite all these benefits and while around two thirds of web pages are likely to be satisfactorily usable, it is currently hard to recommend browsing the web with \javascript{} disabled for all websites, as it would still significantly reduce the number of websites the user could satisfactorily browse, besides the fact that the user may be required to use some websites (e.g., work-related or government websites) that do require \javascript{}.

Browser extensions such as \ublockorigin{}~\cite{ublockorigin}, \noscripext{}~\cite{noscript-extension} and the unmaintained \umatrix{}~\cite{umatrix}, allow users to selectively enable or disable \javascript{} for each domain (and even in a more fine-grained way for some of them), but they require manual action for each visited site, technical knowledge from the user and may not be easily usable on mobile because of reduced screen size.

\subsection{Website Usage of \javascript{} Features With No Fallback}
\subsubsection{Negative Impact of \acs{ui} Frameworks}
\paragraph{Minimum Developer Boilerplate}
The fact that some interactive components are not usable without \javascript{} is often due to them not being written from scratch specifically for the page where they will be used, but instead coming from a \acs{ui} framework, be it an open-source or an in-house one.
To make them easy to use, these \acs{ui} frameworks rely on a set of \acs{css} classes to apply on elements to style and define their behavior.
For instance, \bootstrap{} only requires a couple of classes (\cssclass{dropdown-toggle} and \cssclass{dropdown-menu}) and a few attributes to be added to a button and a list so that they behave as a dropdown list~\cite{bootstrap-dropdowns}.
Unlike the implementation from \autoref{lst:accordion-nojs}, there is no need for the developer to manually insert an extra \htmltag{input} to keep state, since the toggle behavior is entirely handled in \javascript{}, thus requiring minimum manual boilerplate.
\bootstrap{} explicitly documents that its components do not fall back gracefully without \javascript{}, and leaves the implementation of noscript
fallbacks to the developer~\cite{bootstrap-javascript}, only hinting at displaying a noscript warning to tell the user that \javascript{} is required.
The same strategy is followed by other \acs{ui} frameworks, such as \zurbfoundation{}~\cite{foundation-docs} or \semanticui{}~\cite{semanticui-docs}.

\paragraph{Front-End Frameworks}
This trend is exacerbated by the use of front-end frameworks, which require client-side \javascript{} to deliver a significant part of the user interface.
When using frameworks such as \react{}~\cite{react-homepage}, \vuejs{}~\cite{vue-homepage}, \angularjs{}~\cite{angular-homepage}, or \svelte{}~\cite{svelte-homepage}, which all, by default, rely on client-side \javascript{} to build interface components, it may be tempting not to provide any fallbacks since the website is very likely to be significantly broken anyway, regardless of best practices, especially Progressive Enhancement~\cite{webdevnojs}, which recommends separating page semantics from interactivity, while making the former as robust and accessible as possible.

\paragraph{\acl{ssr}/\acl{ssg}}
Some of these frameworks can be used as part of an \acs{ssr} stack, such as \nextjs{}~\cite{nextjs} (for \react{}) or \nuxtjs{}~\cite{nuxtjs} (for \vuejs{}), able to render the page on the server, before sending it to the client, sparing it from the rendering burden and dependency on \javascript{} for rendering the components.
\acs{ssg} can also sometimes be used to render pages ahead of time, when they do not depend on user data, thus reducing the server load.
However, this plays no role in providing client-side fallbacks for interactive elements, such as dropdowns, accordions, or forms.

\subsubsection{\acl{seo} Motivation}
A point of interest for websites to provide \javascript{} fallbacks, at least for basic content, is to improve their ranking on search engines.
Because it is an expensive operation, many search engines and social media crawlers do not execute \javascript{} at all~\cite{searchengine-js-nonindexing,deepcrawl-searchengine-js-nonindexing}, possibly completely missing the page content if it is not rendered without
\javascript{}, thus reducing the chance for the page to be properly indexed by the search engine.
Some of the most popular search engines do run \javascript{}, like \googlesearch{} (since at least~2014)~\cite{googlebotjs2014, googlebotjs2019}.

%% file: crawl_requests_table.tex
\begin{tabular}{lrrrrr}
\toprule
{} & \multicolumn{2}{c}{[plain]} & \multicolumn{2}{c}{[nojs]} & Mean change (\%) \\
{} &       M &   SD &      M & \multicolumn{2}{l}{SD} \\
Request count &         &      &        &      &                                 \\
\midrule
All                          &    72.6 & 59.1 &   28.3 & 34.0 &                           -61.0 \\
--- Non-tracking             &    50.9 & 44.1 &   25.1 & 33.0 &                           -50.7 \\
--- Tracking                 &    21.7 & 29.8 &    3.3 &  8.1 &                           -85.0 \\
First party                  &    27.7 & 31.3 &   16.4 & 24.5 &                           -40.8 \\
--- Image                    &    13.3 & 21.7 &   12.3 & 22.2 &                            -7.6 \\
--- Stylesheet               &     2.6 &  5.0 &    2.4 &  4.6 &                            -6.2 \\
--- Font                     &     1.5 &  2.6 &    1.3 &  2.4 &                           -10.2 \\
--- Script                   &     7.1 & 11.4 &    0.0 &  0.0 &                          -100.0 \\
--- XHR                      &     2.5 &  5.7 &    0.1 &  0.7 &                           -97.6 \\
Third party                  &    44.9 & 49.6 &   11.9 & 26.4 &                           -73.4 \\
--- Image                    &    15.8 & 26.4 &    8.5 & 24.7 &                           -46.3 \\
--- Stylesheet               &     2.1 &  3.6 &    1.4 &  3.5 &                           -30.5 \\
--- Font                     &     2.3 &  3.7 &    1.4 &  2.6 &                           -38.2 \\
--- Script                   &    16.8 & 19.7 &    0.0 &  0.0 &                          -100.0 \\
--- XHR                      &     5.5 &  8.6 &    0.0 &  0.2 &                           -99.8 \\
\bottomrule
\end{tabular}

%% file: conclusion.tex
\section{Conclusion}
In this paper, we performed a crawl on \SI{6,384}{pages} and quantified the use and reliance on \javascript{} of websites to provide content and features that the user was likely to expect when reaching the page.
We found that \percallworking{} of pages were very likely to be completely working with \javascript{} disabled and that more than \percmainworking{} were likely to be usable enough, when potentially broken elements were not part of the main section.
We also observed that reliance on \javascript{} was dependent on the website category, and that it could be really low for some categories, that do not rely much on interactive features.
We finally detailed reasons for why it would be beneficial for websites to be non-\javascript{} friendly and focused on possible reasons for which websites may not currently be supporting non-\javascript{} users.
Based on these findings, we suggest that future work explores solutions to ease adoption of non-\javascript{} friendly implementations of interface features, when it is possible, in accordance with Progressive Enhancement principles.

\section*{Availability}
We make available the complete crawl infrastructure and all breakage detection heuristics at\\\url{https://archive.softwareheritage.org/browse/origin/https://gitlab.inria.fr/Spirals/breaking-bad.git}.

\begin{acks}
This project is funded by the Hauts-de-France region in the context of the ASCOT project of the STaRS framework.
\end{acks}

%% file: appendix.tex
\newpage
\section*{Appendix}

\input{html_elements_js.tex}

\input{ui_framework_js.tex}

\begin{figure}[b]
  \includegraphics[width=\linewidth]{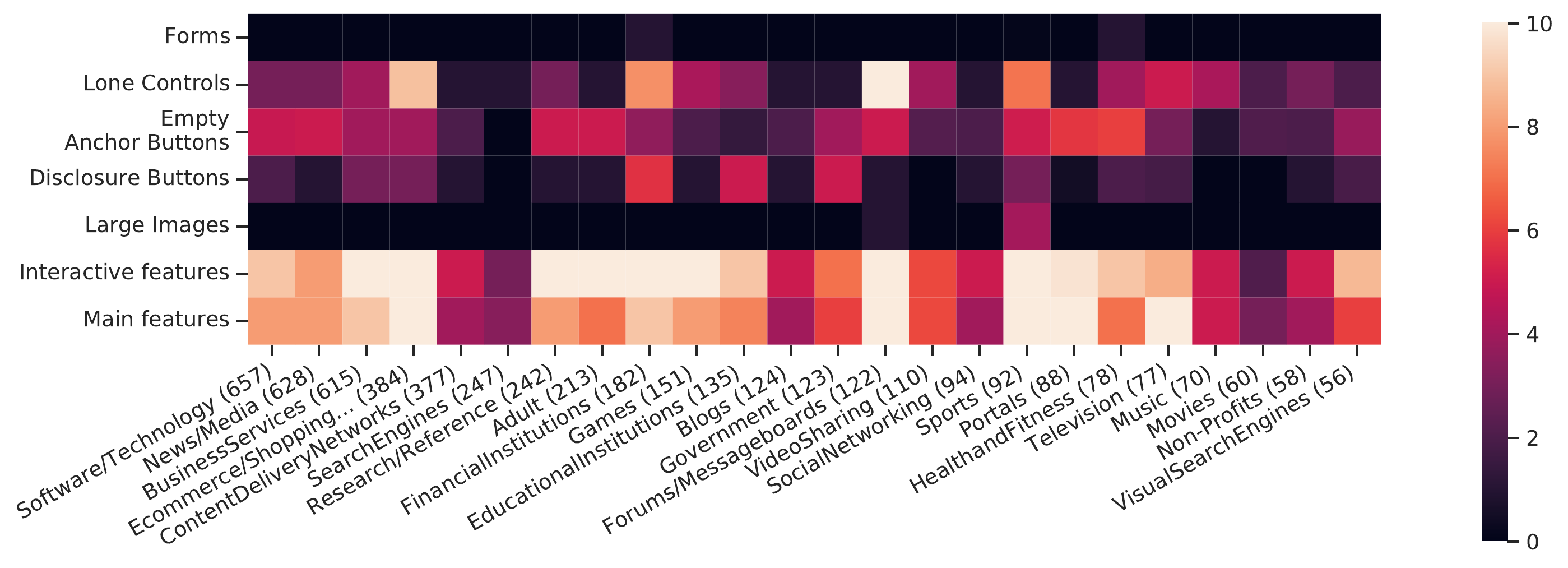}
  \Description{Heatmap of differential breakage of visible elements in the whole page, across website categories. Blogs are much less broken than e-commerce sites when blocking javascript.}
  \caption{Color represents the 90\textsuperscript{th} percentile of differential
    breakage of visible elements in the \textit{whole page}.
  Lighter shades denote higher differential breakage.
  Only categories with more than \catthreshold{} in the dataset are plotted, to
  improve readability.
  This highlights the disparity of breakage across
website categories.}\label{fig:features-all-dist}
\end{figure}

%% file: html_elements_js.tex
\begin{longtable}{p{4cm} p{1.2cm} p{5cm}}
  \toprule
  Element & JavaScript reliance & JavaScript use cases\\
  \midrule\endhead
  \caption{JavaScript reliance of standard \acs{html}
  elements~\cite{mdn-html-elements}.\\
  (0) does not require JavaScript based on the standard, and is very unlikely to
  require JavaScript in the wild,\\
  (1) does not require JavaScript based on the standard, but
  sometimes legitimately uses JavaScript in the wild, for additional features,\\
  (1†) does not require JavaScript based on the standard, but
  sometimes uses JavaScript in the wild, in a non-semantic manner,\\
  (2) does not require JavaScript based on the standard, but
  requires JavaScript in some use cases,\\
  (2*) does not require JavaScript based on the standard, but
  often requires JavaScript when used outside a form,\\
  (3) always requires JavaScript based on the
standard.}\label{tab:js-html-elements}\\
  \toprule
  Element  & JavaScript reliance & JavaScript use cases\\
  \midrule\endfirsthead
  \emph{Main root}\\
  \midrule
  \texttt{<html>} & (0) & \\

  \midrule
  \emph{Document metadata}\\
  \midrule
  \texttt{<base>} & (0) & \\
  \texttt{<head>} & (0) & \\
  \texttt{<link>} & (1) & Asynchronous loading \\
  \texttt{<meta>} & (0) & \\
  \texttt{<style>} & (0) & \\
  \texttt{<title>} & (0) & \\

  \midrule
  \emph{Sectioning root}\\
  \midrule
  \texttt{<body>} & (0) & \\

  \midrule
  \emph{Content sectioning}\\
  \midrule
  \texttt{<address>} & (0) & \\
  \texttt{<article>} & (0) & \\
  \texttt{<aside>} & (0) & \\
  \texttt{<footer>} & (0) & \\
  \texttt{<header>} & (0) & \\
  \texttt{<h1>}, \texttt{<h2>}, \texttt{<h3>}, \texttt{<h4>}, \texttt{<h5>}, \texttt{<h6>} & (0) & \\
  \texttt{<main>} & (0) & \\
  \texttt{<nav>} & (0) & \\
  \texttt{<section>} & (0) & \\

  \midrule
  \emph{Text content}\\
  \midrule
  \texttt{<blockquote>} & (0) & \\
  \texttt{<dd>} & (0) & \\
  \texttt{<div>} & (1†) & Non-semantic button\\
  \texttt{<dl>} & (0) & \\
  \texttt{<dt>} & (0) & \\
  \texttt{<figcaption>} & (0) & \\
  \texttt{<figure>} & (0) & \\
  \texttt{<hr>} & (0) & \\
  \texttt{<li>} & (0) & \\
  \texttt{<ol>} & (0) & \\
  \texttt{<p>} & (0) & \\
  \texttt{<pre>} & (0) & \\
  \texttt{<ul>} & (0) & \\

  \midrule
  \emph{Table content}\\
  \midrule
  \texttt{<caption>} & (0) & \\
  \texttt{<col>} & (0) & \\
  \texttt{<colgroup>} & (0) & \\
  \texttt{<table>} & (0) & \\
  \texttt{<tbody>} & (0) & \\
  \texttt{<td>} & (0) & \\
  \texttt{<tfoot>} & (0) & \\
  \texttt{<th>} & (0) & \\
  \texttt{<thead>} & (0) & \\
  \texttt{<tr>} & (0) & \\

  \midrule
  \emph{Demarcating edits}\\
  \midrule
  \texttt{<del>} & (0) & \\
  \texttt{<ins>} & (0) & \\

  \midrule
  \emph{Inline text semantics}\\
  \midrule
  \texttt{<a>} & (1†) & Non-semantic button\\
  \texttt{<abbr>} & (0) & \\
  \texttt{<b>} & (0) & \\
  \texttt{<bdi>} & (0) & \\
  \texttt{<bdo>} & (0) & \\
  \texttt{<br>} & (0) & \\
  \texttt{<cite>} & (0) & \\
  \texttt{<code>} & (0) & \\
  \texttt{<data>} & (0) & \\
  \texttt{<dfn>} & (0) & \\
  \texttt{<em>} & (0) & \\
  \texttt{<i>} & (0) & \\
  \texttt{<kbd>} & (0) & \\
  \texttt{<mark>} & (0) & \\
  \texttt{<q>} & (0) & \\
  \texttt{<rp>} & (0) & \\
  \texttt{<rt>} & (0) & \\
  \texttt{<ruby>} & (0) & \\
  \texttt{<s>} & (0) & \\
  \texttt{<samp>} & (0) & \\
  \texttt{<small>} & (0) & \\
  \texttt{<span>} & (1†) & Non-semantic button\\
  \texttt{<strong>} & (0) & \\
  \texttt{<sub>} & (0) & \\
  \texttt{<sup>} & (0) & \\
  \texttt{<time>} & (0) & \\
  \texttt{<u>} & (0) & \\
  \texttt{<var>} & (0) & \\
  \texttt{<wbr>} & (0) & \\

  \midrule
  \emph{Image and multimedia}\\
  \midrule
  \texttt{<area>} & (0) & \\
  \texttt{<audio>} & (1) & Custom controls\\
  \texttt{<img>} & (1) & Lazy-loading \\
  \texttt{<map>} & (0) & \\
  \texttt{<track>} & (0) & \\
  \texttt{<video>} & (1) & Custom controls\\

  \midrule
  \emph{Embedded content}\\
  \midrule
  \texttt{<embed>} & (0) & \\
  \texttt{<iframe>} & (0) & \\
  \texttt{<object>} & (0) & \\
  \texttt{<param>} & (0) & \\
  \texttt{<picture>} & (0) & \\
  \texttt{<portal>} & (0) & \\
  \texttt{<source>} & (1) & Lazy-loading \\

  \midrule
  \emph{SVG and MathML}\\
  \midrule
  \texttt{<svg>} & (1) & Embedded JavaScript\\
  \texttt{<math>} & (0) & \\

  \midrule
  \emph{Scripting}\\
  \midrule
  \texttt{<canvas>} & (3) & \\
  \texttt{<noscript>} & (0) & \\
  \texttt{<script>} & (2) & May embed a script or a data block\\

  \midrule
  \emph{Forms}\\
  \midrule
  \texttt{<button>} & (2*) & \\
  \texttt{<datalist>} & (2*) & \\
  \texttt{<fieldset>} & (0) & \\
  \texttt{<form>} & (2) & Requires JavaScript when the form or its values cannot be submitted\\
  \texttt{<input>} & (2*) & \\
  \texttt{<label>} & (0) & \\
  \texttt{<legend>} & (0) & \\
  \texttt{<meter>} & (2*) & \\
  \texttt{<optgroup>} & (2*) & \\
  \texttt{<option>} & (2*) & \\
  \texttt{<progress>} & (2*) & \\
  \texttt{<select>} & (2*) & \\
  \texttt{<textarea>} & (2*) & \\
  \bottomrule
\end{longtable}

%% file: ui_framework_js.tex
\begin{longtable}{p{3.5cm} r r p{1.7cm} r}
  \toprule
  Component
    & Bootstrap~5
    & Foundation~6
    & Tailwind Elements
    & Semantic~UI\\
  \midrule\endhead
  \caption{JavaScript reliance of common UI framework
  components, based on their documentations and manual testing.\\
  (0) does not require JavaScript,\\
  (0*) does not require JavaScript in the documentation, but is likely to be
  used with JavaScript in the wild,\\
  (1) does not require JavaScript to be displayed but requires JavaScript to be
  dismissed, or is used to display transient state, mainly useful with
  JavaScript,\\
  (1*) does not require JavaScript based on the standard, but requires JavaScript in
  some use cases (mostly when used outside a form),\\
  (2) does not require JavaScript to be displayed, but requires JavaScript for
  interactive behavior,\\
  (3) requires JavaScript and displays nothing
otherwise.}\label{tab:js-framework-components}\\
  \toprule
  Component
    & Bootstrap~5~\cite{bootstrap-js-dependency}
    & Foundation~6~\cite{foundation-docs}
    & Tailwind Elements~\cite{tailwind-elements-docs}
    & Semantic~UI~\cite{semanticui-docs}\\
  \midrule\endfirsthead
  Accordion                                  & (2)     & (2)      & (2)       & (2)\\
  Advertisement                              & \na{}   & \na{}    & \na{}     & (0)\\
  Alerts/Message/Callout                     & (1)     & (1*)     & (1)       & (1)\\
  Badge(s)                                   & (0)     & (0)      & (0)/(1)   & \na{}\\
  Breadcrumb(s)                              & (0)     & (0)      & (0)       & (0)\\
  Button(s)                                  & (1*)    & \na{}    & (1*)      & (1*)\\
  Button group                               & (0)     & \na{}    & (0)       & (0)\\
  Card(s)                                    & (0)     & (0)      & (0)       & (0)\\
  Carousel/Orbit                             & (2)     & (2)      & (0*)      & \na{}\\
  Charts                                     & \na{}   & \na{}    & (3)       & \na{}\\
  Chips                                      & \na{}   & \na{}    & (1)       & \na{}\\
  Checkbox/Checks                            & (1*)    & \na{}    & \na{}     & (1*)\\
  Close button                               & (2)     & (2)      & \na{}     & \na{}\\
  Collapse                                   & (2)     & \na{}    & \na{}     & \na{}\\
  Comment                                    & \na{}   & \na{}    & \na{}     & (1*)\\
  Container                                  & \na{}   & \na{}    & \na{}     & (0)\\
  Datepicker                                 & \na{}   & \na{}    & (3)       & \na{}\\
  Dimmer                                     & \na{}   & \na{}    & \na{}     & (1)\\
  Divider                                    & \na{}   & \na{}    & \na{}     & (0)\\
  Drilldow menu                              & \na{}   & (2)      & \na{}     & \na{}\\
  Dropdown(s)                                & (2)     & (2)      & (2)       & (2)\\
  Embed                                      & \na{}   & \na{}    & \na{}     & (0)\\
  Feed                                       & \na{}   & \na{}    & \na{}     & (0)\\
  File input                                 & \na{}   & \na{}    & (0)       & \na{}\\
  Flag                                       & \na{}   & \na{}    & \na{}     & (0)\\
  Floating labels                            & (0)     & \na{}    & \na{}     & \na{}\\
  Footer                                     & \na{}   & \na{}    & (0)       & \na{}\\
  Form validation                            & (0)/(2) & \na{}    & (0)       & (3)\\
  Input group, Layout/Form(s)                & \na{}   & (0)      & (1*)      & (1*)\\
  Gallery                                    & \na{}   & \na{}    & (0)       & \na{}\\
  Grid                                       & \na{}   & \na{}    & \na{}     & (0)\\
  Headings/Header                            & \na{}   & \na{}    & (0)       & (0)\\
  Image(s)                                   & \na{}   & \na{}    & (0)       & (0)\\
  Icon                                       & \na{}   & \na{}    & \na{}     & (0)\\
  Item                                       & \na{}   & \na{}    & \na{}     & (0)\\
  Form controls/Input(s)                     & \na{}   & \na{}    & (1*)      & (1*)\\
  Label                                      & \na{}   & (0)      & \na{}     & (0)\\
  List group/List                            & (0)     & \na{}    & (0)       & (0)\\
  Menu                                       & \na{}   & (0)      & \na{}     & (0)\\
  Media                                      & \na{}   & (0)      & \na{}     & \na{}\\
  Modal/Reveal                               & (2)     & (2)      & (2)       & (2)\\
  Multiselect                                & \na{}   & \na{}    & (1*)      & \na{}\\
  Navs \& tabs/Tab(s)/Pills                  & (2)     & (2)      & (2)       & (2)\\
  Navbar/Topbar                              & (0)     & (0)      & (0)       & \na{}\\
  Offcanvas/Sidebar                          & (2)     & (2)      & \na{}     & (2)\\
  Pagination                                 & (0)     & (0)      & (0)       & \na{}\\
  Placeholder(s)                             & (0)     & \na{}    & \na{}     & (1)\\
  Popover(s)/Popup                           & (2)     & \na{}    & (2)       & (2)\\
  Progress/Progress Bar                      & (1)     & (1)      & (1)       & (3)\\
  Radios                                     & (1*)    & \na{}    & (1*)      & \na{}\\
  Rail                                       & \na{}   & \na{}    & \na{}     & (0)\\
  Rating                                     & \na{}   & \na{}    & (0)       & (3)\\
  Range/Slider                               & (1*)    & (2)      & (2)       & \na{}\\
  Responsive Accordion Tabs                  & \na{}   & (2)      & \na{}     & \na{}\\
  Responsive Embed                           & \na{}   & (0)      & \na{}     & \na{}\\
  Responsive Navigation                      & \na{}   & (2)      & \na{}     & \na{}\\
  Reveal                                     & \na{}   & \na{}    & \na{}     & (0)\\
  Scrollspy/Magellan                         & (2)     & (2)      & \na{}     & \na{}\\
  Search(s)                                  & \na{}   & \na{}    & (1*)      & \na{}\\
  Select                                     & (1*)    & \na{}    & (1*)      & \na{}\\
  Segment                                    & \na{}   & \na{}    & \na{}     & (0)/(1)\\
  Shape                                      & \na{}   & \na{}    & \na{}     & (3)\\
  Sidenav                                    & \na{}   & \na{}    & (0)       & \na{}\\
  Spinners/Loader                            & (1)     & \na{}    & (1)       & (1)\\
  Statistic                                  & \na{}   & \na{}    & \na{}     & (0)\\
  Sticky                                     & \na{}   & \na{}    & \na{}     & (2)\\
  Switch                                     & \na{}   & (1*)     & (1*)/(2)  & \na{}\\
  Stepper/Step                               & \na{}   & \na{}    & (0)       & (0)\\
  Table(s)                                   & \na{}   & (0)      & (0)       & (0)\\
  Thumbnail                                  & \na{}   & (0)      & \na{}     & \na{}\\
  Textarea                                   & \na{}   & \na{}    & (1*)      & \na{}\\
  Timepicker                                 & \na{}   & \na{}    & (1*)      & \na{}\\
  Toast(s)                                   & (1)     & \na{}    & (1)       & \na{}\\
  Tooltip(s)                                 & (2)     & (2)      & (2)       & \na{}\\

  \bottomrule
\end{longtable}